\documentclass[aps,prx,twocolumn]{revtex4-2}	
\usepackage{graphicx}
\usepackage{amsmath, amsfonts,amssymb}
\usepackage{color}
\usepackage[dvipsnames]{xcolor}

\newcommand{\be}{\begin{equation}}
\newcommand{\ee}{\end{equation}}

\newcommand{\bea}{\begin{eqnarray}}
\newcommand{\eea}{\end{eqnarray}}

\newcommand{\Tr}{{\rm Tr}}

\usepackage{outlines}

\usepackage{mathrsfs}

\renewcommand{\vec}[1]{{\bf #1}}

\renewcommand{\mod}{{\rm mod}\ }

\newcommand{\norm}[1]{\left\lVert#1\right\rVert}
\definecolor{MyColor}{RGB}{0,0,240}
\renewcommand{\vec}[1]{{\bf #1}}

\newcommand{\llangle}{\langle \!\langle}
\newcommand{\rrangle}{\rangle \! \rangle}

\newcommand{\T}{{\tilde{T}}}
\newcommand{\Ot}{{\tilde{\Omega}}}
\newcommand{\Heff}{\mathcal{H}_{\rm eff}}

\begin{document}
\title{
Quantum frequency locking and  down-conversion in a driven qubit-cavity system
}
\author{Frederik Nathan$^1$, Gil Refael$^2$, Mark S. Rudner$^1$, and Ivar Martin$^3$}
\affiliation{$^1$Center for Quantum Devices, Niels Bohr Institute,  University of Copenhagen, 2100 Copenhagen, Denmark \\
$^2$Institute for Quantum Information and Matter, Caltech, Pasadena, California 91125, USA\\
$^3$ Materials Science Division, Argonne National Laboratory, Argonne, Illinois 60439, USA
}
\begin{abstract}

We study a periodically driven qubit coupled to a quantized cavity mode. 
Despite its apparent simplicity, this system supports a rich variety of exotic phenomena, such as topological frequency conversion as recently discovered in [Martin {\it et al}, PRX {\bf 7}, 041008 (2017)]. 
Here we report on a qualitatively different phenomenon that occurs in this platform,  where the cavity mode's oscillations lock their frequency to  a rational fraction $r/q$ of the driving frequency $\Omega$.
This phenomenon, which we term quantum frequency locking,  is characterized by the emergence of  $q$-tuplets of stationary (Floquet) states whose quasienergies are separated by  $\Omega/q$, up to exponentially small corrections.
The Wigner functions of these states are nearly identical, and exhibit highly-regular and symmetric  structure in phase space.
Similarly to Floquet time crystals, these  states 
underlie 
discrete time-translation symmetry breaking in the model.
We develop a semiclassical  approach for analyzing and predicting quantum frequency locking in the model, and use it to identify the conditions under which it occurs. \end{abstract}
\date{\today}
\maketitle

\section{Introduction}
In recent years, periodic driving has been explored % extensively
as a way to create desirable properties in 
otherwise ordinary systems~\cite{Dalibard_2011,Eckardt_2017,Cooper_2019,Oka_2019_review,Rudner_2019_review,Harper_2019_review}.
In addition to inducing exotic phases that already exist in equilibrium~\cite{Oka_2009,Lindner_2011,Jotzu_2014}, periodic driving can also induce exotic phenomena with no equilibrium counterpart~\cite{Kitagawa2010, Jiang2011,Rudner_2013, Titum_2016, Khemani_2016, Else_2016b, Martin_2017,Nathan_2019,Crowley, Kolodrubetz_2018, Nathan_2019b, ChandranSondhi,Fidkowski,Roy,Fidkowski2,TC-potter,Else2,Keeling,Gong,sondhi1,Huang_2018,Efetov,Peer,Mathey, Yao-Demler, Ueda, AMRey}.
These predictions inspired a wide range of experiments, leading to the realization and observation of
 new drive-induced phenomena, such as Floquet time crystals, and anomalous Floquet insulators~\cite{Broome2010,Sacha_2015,Hu2015,vonKeyserlingk_2016,Choi_2017, Zhang_2017, Maczewsky2017, Mukherjee2017, Moessner_2017,Ho_2017,Russomano_2017,McIver_2019,Lukin1,Wintersperger2020}.  
%%%%%%%%%%%%%%%%%%%%%%%%%%%%%%%%%%%%%%%%%%%%%%%%%%%%%%%%%%%%%%
\begin{figure}[!ht]
\includegraphics[width=0.99\columnwidth]{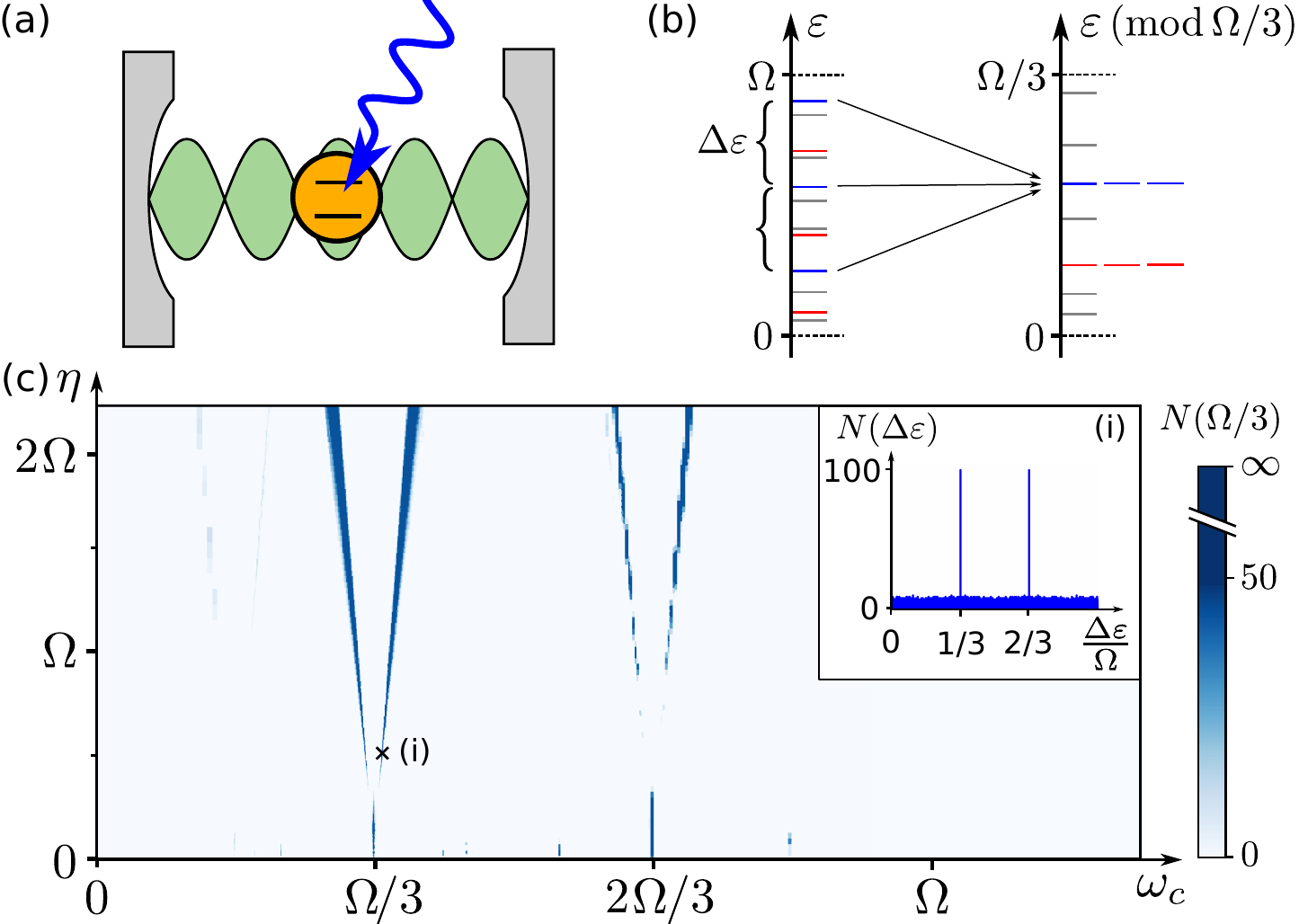}
\caption{(a) 
We study a  2-level system (orange), coupled to a single cavity mode (green) while driven periodically with  frequency $\Omega$ (blue).
(b) Quasienergy locking: while the quasienergies of the system (gray, red, and blue lines)
are effectively uniformly distributed between $0$ and $\Omega$, in certain finite parameter regimes 
the spacings between certain subsets of levels (red and blue)  are exponentially close to  
$\Omega /q$ for some integer $q$
(here we illustrate this clustering for $q=3$).
(c) Number of period-3 quasienergy-locked levels %Floquet eigenstates
in the model, obtained from numerical exact evolution and diagonalization of the system's Floquet operator, as function of cavity frequency $\omega_{\rm c}$ and qubit-cavity coupling $\eta$ (see Sec.~\ref{sec:model} for model details, Sec.~\ref{sec:numerics} for details of the simulation, {and Appendix~\ref{app:higher_mode_locking} for similar plots for other frequency locking ratios)}.
%{\it Inset:} 
The inset shows a histogram of quasienergy level spacings in the frequency-locked regime (parameters indicated by the cross  in main panel).
}
\label{fig:Fig1}
\end{figure}
%%%%%%%%%%%%%%%%%%%%%%%%%%%%%%%%%%%%%%%%%%%%%%%%%%%%%%%%%%%%%%

In this work, we consider another class of such %novel
driving-induced phenomena:  quantum frequency locking~\cite{Holthaus_1994,Sacha_2015,Zhang_2017,Sacha_2017,Huang_2018,Giergiel_2018,Pizzi_2019,Matus_2019}.
Quantum frequency locking arises when a quantum system with an intrinsic characteristic frequency $\omega_{\rm c}$ is driven at a frequency $\Omega = 2\pi/T$ close to (but not necessarily equal to) a rational multiple $q/r$ of the intrinsic frequency $\omega_{\rm c}$. 
In this case, the system can  respond by robust oscillations with period locked {\it exactly}  to an integer  multiple of the driving period,  $q T$. %(2\pi/\Omega)$.

Here we propose a new and accessible realization of quantum frequency locking. Namely, we consider a periodically-driven qubit coupled to a quantized electromagnetic cavity (Fig.~\ref{fig:Fig1}a). 
When the qubit is driven close to resonance with a rational multiple  of the cavity's resonance frequency, the cavity mode oscillates with frequency  locked to $r\Omega/q$. %\addIM{need to check the usage or q/r throughout; there are several flips}.
This phenomenon has a wide range of interesting implications and uses, which we explore in this paper:
in particular,  it implies the formation of characteristic  subsets of quasienergy levels which are separated by % {exactly} 
$\Omega /q$, up to exponentially small corrections (see sketch in Fig.~\ref{fig:Fig1}b)~\cite{Holthaus_1994,Zhang_2017}; we term this related phenomenon ``quasienergy locking.''
% regime where the system oscillates with a  {exactly} to $r\Omega/q$. 
Quantum frequency locking   is, moreover, a robust effect, which does not require  fine-tuning~\cite{frequency_tuning}. It persists both for weak and strong qubit-cavity coupling, and for finite ranges of the driving frequency  (see Fig.~\ref{fig:Fig1}c).

Quantum frequency locking has previously been considered in various platforms and
settings, including ultracold atoms interacting with a vibrating mirror~\cite{Sacha_2015,Matus_2019}, spin chains or clock-models~\cite{Russomano_2017,Surace_2019,Pizzi_2019}, parametrically driven chains of electromagnetic cavities \cite{huang2020non}, and  nonlinear oscillators~\cite{Holthaus_1994,Guo_2013,Zhang_2017}. 
Our % driven spin-cavity
proposal provides a new and complimentary realization that can capitalize on recent advances in control of  few-level quantum systems in Rydberg atoms, quantum dots, and superconducting qubits interacting with microwave cavities~\cite{Gross_2017,Burkard_2020,Clerk_2020}.
In this way our results provide a direct path for realizing quantum frequency locking on readily-available experimental platforms.
In addition to being of fundamental interest, the robust coherent oscillations of the cavity mode can moreover
be used as a frequency converter to generate a coherent signal at a frequency different from the drive (by a factor $r/q$). 

Quantum frequency locking has a well-established classical counterpart, known as Arnold Tongues~\cite{Rasband_1990,strogatz, Peer, Armour, Gabrielse}. 
Quantum mechanics, however, introduces several new aspects to this effect:
the wave-packets of a  system do not spread in phase space when observed stroboscopically~\cite{Buchleitner_2002, huang2020non}.
Moreover,  as explained above, the robust period-multiplied oscillations implies a nontrivial ordering of the quasienergy spectrum in the system (see Fig.~\ref{fig:Fig1}b).
The Wigner functions of the corresponding Floquet eigenstates  exhibit a remarkably rich structure (see Fig.~\ref{fig:contours}cd below).

The nontrivial organization of the quasienergy spectrum mentioned above is a signature of the breakdown of discrete time-translation symmetry~\cite{Holthaus_1994,Sacha_2015,Bruder,Zhang_2017}. 
Thus frequency-locked quantum systems also present examples of ``Floquet time-crystals''~\cite{Khemani_2016,Else_2016b}, %[Time crystal papers], 
and %strengthen the assertion 
demonstrate how time-translation symmetry breaking (in a broader sense than defined in Ref.~\onlinecite{Else_2016b}) may be realized in few-body quantum systems [see also Refs.~\onlinecite{Sacha_2015,Zhang_2017}]; see Sec.~\ref{sec:quasienergy_locking} for further discussion.

In what follows we introduce the qubit-cavity model we study
in Sec.~\ref{sec:model}. 
In Sec.~\ref{sec:theory}, we perform an approximate semiclassical analysis of the model  to describe how   quantum frequency locking  emerges, and identify the conditions under which it occurs. 
We discuss the implications of quantum frequency locking for the quasienergy spectrum of the system in \ref{sec:phase_locking}, before confirming our approach numerically in Sec.~\ref{sec:numerics}. 
Here we also discuss how quantum frequency locking may be utilized for frequency conversion (Sec.~\ref{sec:signatures}).
We conclude with a discussion in Sec.~\ref{sec:discussion}.
Technical details of our analysis are provided in the Appendices.

\section{model}
\label{sec:model}
The system we consider consists of a two-level system, such as a qubit, quantum dot, or a spin-1/2 magnetic moment, coupled linearly to a quantized  electromagnetic cavity mode, and to a periodic drive (see  Fig.~\ref{fig:Fig1}a). 
Without loss of generality, we  refer to the two-level system simply as a spin below.

The Hamiltonian of the  system is given by 
\be 
\hat H(t) = \hat H_{\rm c} 
 +\hat  H_{\rm s}(t).
\label{eq:hamiltonian_1}
\ee
Here $\hat H_{\rm c}$ and $\hat H_{\rm s}(t)$ denote the  Hamiltonian  of the cavity and spin, respectively, defined such that  $\hat H_{\rm s}(t)$      includes the spin-cavity coupling;  this term [and hence also $\hat H(t)$] depends on time $t$.
The cavity Hamiltonian is simply given by $\hat H_{\rm c} = \omega_{\rm c} \hat b^\dagger \hat b$, where $\omega_{\rm c}$ and $\hat b$ denote  the frequency and bosonic annihilation operator of the cavity mode, respectively (here and below, we work in units where $\hbar = 1$). 
The Hamiltonian of the spin, $\hat H_{\rm s}(t)$, consists of three parts: a static (Zeeman) part, $\hat H_0$,   a term coupling the spin to a  time-dependent  driving field, $\hat V_{\rm dr}(t)$, and a term  coupling the spin
to the cavity field, $\hat H_{\rm sc}$: 
\be 
\hat H_{\rm s}(t) = \hat H_0 +\hat  V_{\rm dr}(t) +\hat  H_{\rm sc}.
\ee
The drive encoded in $\hat V_{\rm dr}(t)$ has $T$-periodic time-dependence  (angular frequency $\Omega \equiv{2\pi}/{T}$): $\hat V_{\rm dr}(t)= \hat V_{\rm dr}(t+T)$. 

We do not expect frequency-locking to depend on the specific details of $\hat H_0$, $\hat V_{\rm dr}(t)$ and $\hat H_{\rm sc}$. For concreteness, however, we use the forms:
\begin{eqnarray}
\hat H_0 &=& \eta \hat \sigma_x B_0,\\
\hat V_{\rm dr} (t)&=& \eta A_d [\sin(\Omega t)\hat  \sigma_x + \cos(\Omega t)\hat \sigma_z],\\
\hat H_{\rm sc} &=& \eta ( \hat b \hat \sigma^+ + \hat b^\dagger\hat  \sigma^-)\label{eq:j_sc_def}.
\end{eqnarray}
Here   $\eta$ parametrizes the spin's coupling to %all fields
the external (Zeeman and driving) fields and to the cavity field, $\hat \sigma_x,\hat \sigma_y,\hat \sigma_z$   denote the Pauli matrices acting on the  spin, with $\hat \sigma^{\pm} \equiv \frac{1}{2} (\hat \sigma_x \pm i \hat \sigma_y)$, and $B_0$ and $A_d$ are dimensionless numbers denoting  the effective Zeeman field strength and driving amplitude, respectively. This model was shown to support topological frequency conversion  in Refs.~\onlinecite{Martin_2017,Nathan_2019, Crowley}.

The  cavity mode is described by a harmonic oscillator, and can thus %, %being an harmonic oscillator, 
 conveniently be represented using the dimensionless position and momentum operators \cite{x_p_normalization}  $ \hat x \equiv \frac{1}{2} (\hat b+\hat b^\dagger)$ and $ \hat p  \equiv\frac{1}{2i}(\hat b-\hat b^\dagger)$.
In terms of these  operators,  the Hamiltonian of the full system is given by 
\be 
\hat H(t) = \frac{\omega_c}{2}(\hat  x^2+ \hat p^2) + \eta \vec b(\hat  x ,\hat  p ,t)\cdot \vec{\hat  S},
\label{eq:Ht}
\ee
where  $\vec{\hat S} = (\hat \sigma_x,\hat \sigma_y,\hat \sigma_z)$ denotes the  %(pseudo)
{effective} spin operator {describing the qubit}, while % of the two-level system,  while 
\be 
\vec b(x,p,t)= (B_0-A_d\sin\Omega t -x,p,A_d\cos \Omega t )
\label{eq:h_def}
\ee 
can be seen as the effective Zeeman  field acting on the spin, as a function of the cavity mode's position and momentum, and  time. 
%  This quantity will play an important role in the following.

As mentioned in the introduction, the model above can be realized in various ways. Most appealing perhaps are realizations using superconducting qubits~\cite{manucharyan2009fluxonium, PhysRevX.9.041041, Clerk_2020} and nitrogen-vacancy (NV) centers~\cite{nv, PhysRevLett.113.197601},  as well as atoms in optical cavities (see, e.g., Refs.~\onlinecite{Schleier-Smith,Lev-Cavity}).
We expect that our following discussion generalizes to cavities with multiple modes~\cite{Nathan_2019}.

\section{Semiclassical picture of Frequency locking} 
\label{sec:theory}

When the  driving field of the in model of  Sec.~\ref{sec:model} has an off-resonant frequency, $\Omega$, close   to a rational multiple $q/r$ of the cavity eigenfrequency $\omega_{\rm c}$  (where $q$ and $r$ are integers), there exists finite regions of phase space where  the  cavity mode responds to the drive with coherent oscillations whose frequency is  locked {\it exactly} to $r\Omega/q$.
We refer to this phenomenon as {quantum frequency locking}. 
In this section, we demonstrate from heuristic semiclassical arguments how quantum frequency locking arises in the model. 
%presented in Sec.~\ref{sec:model}.
We confirm our approach using numerical simulations in  Sec.~\ref{sec:numerics}. %, and expand its justification analytically in Appendix~\ref{app:}. 

% The approach in Appendix~\ref{app:photon_lattice} may also be used on a wider class of quantum systems than the qubit-cavity system we consider here. }

Our first step towards deriving frequency locking % in the spin-cavity system
is to transform the cavity mode's degrees of freedom, $(\hat x,\hat p)$, to a  frame % where the phase space
rotating with frequency $\tilde \Omega \equiv r\Omega/q$ (as in previous studies of quantum frequency locking; see, e.g., Refs.~\cite{Guo_2013,Zhang_2017}). 
In this rotating frame, the cavity mode becomes much slower than the driving period and the spin's dynamics. 
%We will later
In Sec.~\ref{sec:effective_hamiltonian}, we identify conditions under which this separation of time scales allows us to effectively integrate out the spin and the driving field. %and show how this results in 
This %allows us to obtain 
results in a time-independent semiclassical  effective  Hamiltonian that governs the evolution of the cavity mode in the rotating frame:
\be 
\mathcal H_%{\rm cav}^
{\rm eff}
(x,p) = \frac{\delta \omega}{2} \left(x^2+p^2\right) + \varepsilon (x,p),
\label{eq:heff}
\ee 
where $x$ and $p$ denote the semiclassical position and momentum variables of the cavity mode (in the rotating frame), and $\delta \omega\equiv \omega_c- \tilde \Omega$ is the cavity detuning from $\tilde \Omega$.
The potential $\varepsilon (x,p)$ in Eq. (\ref{eq:heff}) results from  integrating out the spin and the driving field, and plays a central role in our analysis. 
We  identify two distinct parameter regimes where the above separation of timescale occurs, namely,   the large-$\eta$ {\it adiabatic} regime, and the small-$\eta$ {\it Floquet} regime (see Sec.~\ref{sec:effective_hamiltonian} for discussion of these regimes).
Both regimes support quantum frequency locking, but result    in two distinct expressions for the potential $\varepsilon (x,p)$.

Frequency locking can naturally be understood  by %if we
inspecting the effective Hamiltonian $\mathcal H_{\rm eff}(x,p)$:  in Fig.~\ref{fig:contours}ab we plot $\mathcal H_{\rm eff}(x,p)$ 
for the two regimes  where  quantum frequency locking occurs  with period $3$.
In both cases $\mathcal H_{\rm eff}(x,p)$ has local extrema at nonzero values of $x$ and $p$. 
When the  cavity mode is initialized at one of these extrema, its phase space  location  in the rotating frame  remains stationary. 
As a result, in the original ``lab'' frame, the cavity mode will  oscillate with frequency $\Omega/3$.

The picture above also explains the robustness of  frequency locking.
 Even if the cavity is initialized near (but not precisely at) the extremum of $\mathcal H_{\rm eff}$, its location in the rotating frame will remain confined near the extremum at all times. 
Thus,  frequency locking can be achieved  with only moderate requirements for control over initial conditions; for instance, in Fig.~\ref{fig:contours}, it will occur with significant probability as long as the displacement amplitude of the cavity mode, $\sqrt{x^2+p^2}$, is less than $30$ {in the dimensionless units we have adopted [see text above  Eq.~\eqref{eq:Ht}].
The corresponding motion in the lab frame must remain close to this point each time three driving periods have passed. As a result, the frequency spectrum of the cavity mode's motion features a sharp, well-defined peak at $\Omega/3$. 
The extrema moreover cannot be removed by weak perturbations, implying that frequency locking persists in finite parameter ranges.
% We elaborate further on this robustness in Sec.~\ref{sec:phase_locking}.

In Appendix~\ref{app:photon_lattice}  we present a different, complementary perspective on quantum frequency locking, based on the dynamics in the combined Fock space of the oscillator and driving field.
The   approach presented there can in principle be  used to study quantum frequency locking for any driven finite-dimensional quantum system coupled to a quantized cavity mode, in the limit of small nonlinearity and detuning.

 %%%%%%%%%%%%%%%%%%%%%%%%%%%%%%%%%%%%%%%%%%%%%%%%%%%%%%%%%%%%%%%%%%%%%%%%%%%%%%%%%%%%%%%%
\begin{figure}
\includegraphics[width=0.99\columnwidth]{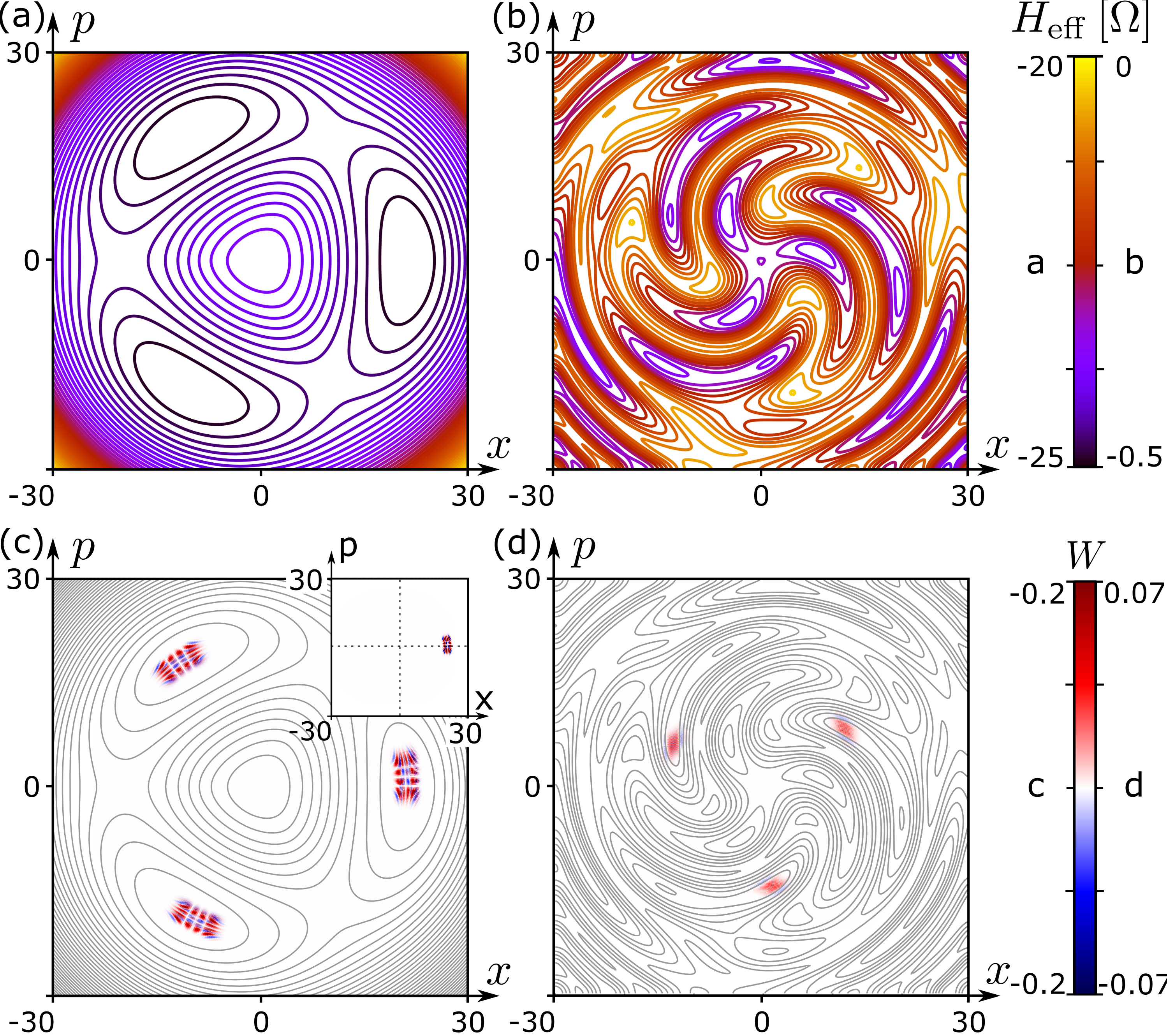}
\caption{(a) and (b): Constant (quasi)energy contours of the effective  Hamiltonian for the cavity mode in the frame where phase space rotates with frequency $\Omega/3$, $\mathcal H_{\rm eff}(x,p)$, [see Eq.~\eqref{eq:heff}],
Panels (a) and (b) depict two parameter sets where period-$3$ quantum frequency locking occurs, within the adiabatic and Floquet regimes, respectively; see Sec.~\ref{sec:effective_hamiltonian} for parameters.
(c) and (d): Wigner functions of quasienergy-locked Floquet eigenstates, using the same parameters as in panels (a) and (b), respectively. 
The contours of the corresponding effective Hamiltonians from panels (a) and (b) shown as grey lines.
{\it Inset, panel} (c):  Wigner function for  an equal-weight linear combination of the $3$ related   Floquet eigenstates of which one is depicted in (c) (see Sec.~\ref{sec:phase_locking}). }
\label{fig:contours}
\end{figure}
%%%%%%%%%%%%%%%%%%%%%%%%%%%%%%%%%%%%%%%%%%%%%%%%%%%%%%%%%%%%%%%%%%%%%%%%%%%%%%%%%%%%%%%%

\subsection{Effective cavity mode  Hamiltonian}
\label{sec:effective_hamiltonian}
We now obtain the effective Hamiltonian in Eq.~\eqref{eq:heff}. % for the two parameter regimes mentioned above. % adiabatic and Floquet regimes. 
To this end,  we consider the dynamics of the system in the rotating frame that was described above. 
The Hamiltonian of the system in this rotating frame is given by $\tilde H(t) = \hat U^\dagger_0(t) \big[\hat H(t)- \tilde \Omega \hat b^\dagger\hat  b\big] \hat U_0(t) ,
\label{eq:rf_transf}
$
where the unitary operator $\hat U_0(t)  \equiv e^{-i \tilde \Omega \hat b^\dagger\hat  b t}$   generates the transformation to the rotating frame:
 the Schr\"odinger equation in the lab frame is solved by $|\psi(t)\rangle = \hat U_0(t)|\tilde \psi(t)\rangle$, where $\partial _t |\tilde \psi(t)\rangle =  -i \tilde H(t)|\tilde \psi(t)\rangle$.
Noting that $\hat U_0(t)$ only acts nontrivially on $\hat H_{\rm sc}$, we find 
\be 
\tilde H(t)   = \frac{\delta \omega}{2} (\hat  x ^2 + \hat  p ^2) + \eta \vec{h}(\hat  x,\hat  p,t) \cdot \vec{\hat S},
\label{eq:rotating_frame_hamiltonian}
\ee
where $\vec h$ is obtained from $\vec b$ in Eqs.~(\ref{eq:Ht}) and (\ref{eq:h_def}) after rotating the oscillator phase space  by $\tilde \Omega t$: 
$\vec h ( x, p,t)= \vec b\big(x \cos\tilde \Omega t +   p \sin\tilde \Omega t, p \cos\tilde \Omega t -   x \sin\tilde \Omega t,t\big)$.
Note that the  Hamiltonian $\tilde H(t)$ in Eq.~\eqref{eq:rotating_frame_hamiltonian} describes a periodically driven system with {extended} period $\tilde T = qT$, through the explicit time-dependence of $\vec h(x,p,t)$ above (recall that $\tilde{\Omega} = r\Omega/q$).

To derive the effective semiclassical Hamiltonian for the cavity mode,
$\mathcal H_{\rm eff}(x,p)$, we consider the  equations of motion generated by $\tilde H(t)$ for the Heisenberg picture operators  $\hat x(t)$, %: % 
$\hat p(t)$, and $\vec{\hat S}(t)$:
\begin{align}
\partial _t \hat x(t)&= \delta \omega \, \hat p(t)+ \eta \vec v_p(t) 
\cdot \vec{\hat S},
\label{eq:x_eom}\\ 
\partial _t \hat p(t) &=  -\delta \omega\, \hat x(t) - \eta  \vec v_x(t) \cdot \vec{\hat S} ,
\label{eq:p_eom}\\
\partial _t{\vec{\hat S}}(t) &=  \eta \vec h[\hat x(t),\hat p(t),t] \times \vec {\hat S}.
\label{eq:s_eom}
\end{align}
where $\vec v_x(t)$ and $\vec v_p(t)$ are  vectors with unit norm: 
$\vec v_x(t)\equiv (-\cos\tilde\Omega t,\sin\tilde \Omega t,0)$,  $\vec v_p(t)\equiv (\sin\Ot t,\cos \Ot t,0)$.
We apply a semiclassical approximation to the equations of motion above, by assuming that the cavity mode's location in phase space is relatively well-defined at all times: 
in Eq.~\eqref{eq:s_eom}, we  approximate $\vec h(\hat x(t) , \hat p(t) , t ) \approx \vec h(x(t),p(t),t)$, where $x(t)\equiv \langle \hat x(t)\rangle$, and $p(t)\equiv \langle \hat p(t)\rangle$ denote the expectation values of the position and momentum operators.
We expect this % semiclassical 
approximation to be justified when the characteristic  scales in phase space of %associated with 
variations in $\vec h(x,p,t)$ are   large compared to the scale of quantum fluctuations $\Delta x,\Delta p \sim 1$. 

The approximation above reduces  Eq.~\eqref{eq:s_eom} to a Bloch-equation with a time-dependent field $\vec h(x(t),p(t),t)$. 
By moreover taking the expectation values on both sides of Eqs.~\eqref{eq:x_eom}-\eqref{eq:s_eom},  we then obtain three coupled  equations of motion for the (semi)classical variables $x(t),p(t),$ and $\vec S(t)\equiv \langle \vec S(t)\rangle$: namely Eqs.~\eqref{eq:x_eom}-\eqref{eq:s_eom} with the operators $\hat x(t)$, $\hat p(t) $, and $\vec{\hat S}(t)$  replaced by their expectation values  $ x(t)$, $p(t) $, and $\vec{ S}(t)$. 
These equations of motion are generated by the time-dependent classical Hamiltonian $\tilde {\mathcal H}(t)$, given by 
%obtained by substituting $\hat{\vec S}$, $\hat x$, and $\hat p$ with their classical counterparts in Eq.~\eqref{eq:rotating_frame_hamiltonian}:
\be 
\tilde{\mathcal H}(t)   = \frac{\delta \omega}{2} (  x ^2 +   p ^2) + \eta \vec{h}(  x,  p,t) \cdot \vec{ S}.
\label{eq:classical_rotating_frame_hamiltonian}
\ee
Note that the dynamics of the cavity mode, $(x(t),p(t))$, has characteristic frequencies $\delta \omega, \eta$, which can be much smaller than $\tilde\Omega$ (which is %near 
on the same order as the resonance frequency of the oscillator) and the characteristic frequencies of the spin's dynamics. % of the spin. 
Below, we identify two parameter regimes where this separation of timescales allows us to effectively eliminate the spin, $\vec S(t)$, and obtain the static effective Hamiltonian for the cavity mode in Eq.~\eqref{eq:heff}.

\subsubsection{Adiabatic regime}
\label{sec:adiabatic}
The simplest ``adiabatic'' regime occurs for large $\eta$. In this regime  the direction of the instantaneous Zeeman field, $\vec h(x(t),p(t),t)$, changes adiabatically with respect to the (fast) Larmor precession of the spin, which has frequency $\sim\eta |\vec h(x,p,t)|$ [see Eq.~\eqref{eq:classical_rotating_frame_hamiltonian}]. 
In this case, the equations of motion [Eq.~\eqref{eq:s_eom} with $\hat x, \hat p$, and $\vec{\hat S}$ substituted by their expectation values $x$ and $p$] has the two distinct solutions: $\vec S(t) \approx \pm \vec h(x,p,t)/|\vec h(x,p,t)|$~\cite{spin_initialization}.
With these solutions for $\vec S$, Eq.~\eqref{eq:classical_rotating_frame_hamiltonian} becomes 
\begin{equation}
\tilde{\mathcal{H}}^{\pm}_{\rm cav}(x,p,t) = \frac{\delta\omega }{2} (x^2 + p^2)  \pm \eta  |\vec h(x,p,t)|.\label{heffa}
\end{equation}

Similarly to quantum systems, the stroboscopic time-evolution generated by %the time-dependent classical Hamiltonian
$\tilde {\mathcal H}^{\pm}_{\rm cav}(x,p,t)$   (i.e.,  time-evolution at integer multiples of $\tilde T$) is equivalent to that generated by some time-independent effective classical Hamiltonian~\cite{Oteo_1991}.
When the cavity mode in the rotating frame oscillates  slowly compared to %the driving frequency 
$\tilde \Omega$ (see below for more detailed conditions), 
%a Magnus-like expansion can be used to show that 
this effective Hamiltonian %$\mathcal H_0(x,p)$ %this effective Hamiltonian 
is well-approximated  by the time-average of $\tilde{\mathcal H}_{\rm cav}^{\pm}(x,p,t)$; i.e., by $\mathcal H_{\rm eff}(x,p)$ in Eq.~\eqref{eq:heff}, with 
\be 
\varepsilon (x,p) =  \pm \frac{\eta}{\T} \int_0^{\T}\!\!\! {\rm d}t\, |\vec h(x,p,t)|, %\varepsilon (x,p)
\label{eq:heff_expression_adiabatic}
\ee
where the sign ($\pm$) depends on the initial alignment of the spin~\cite{spin_initialization}.
Eq.~\eqref{eq:heff_expression_adiabatic} can be obtained using a Magnus expansion of the evolution operator generated by the system's Liouvillian (see Ref.~\cite{Oteo_1991} for details).

The considerations above show that $\mathcal H_{\rm eff} (x,p)$  in Eqs.~\eqref{eq:heff}~and~(\ref{eq:heff_expression_adiabatic})  describes the system when the Larmor precession frequency of the spin, $\eta|\vec h(x,p,t)|$ is much larger than the driving frequency, which in turn should be much larger than the characteristic frequency of the cavity mode in the rotating frame. %(since both the driving period and the cavity mode's dynamics should be slow compared to the Larmor frequency for the spin dynamics to be adiabatic). 
The latter is given by the {\it renormalized} frequency detuning $\delta \omega'(x,p)$, given  by the radial gradient of $\mathcal H_{\rm eff} (x,p)$, divided by the amplitude of the cavity mode, $A_{\rm cav} = \sqrt{x^2 + p^2}$. 
Hence the adiabatic regime arises when $\delta \omega'(x,p) \ll \Omega\ll\eta|\vec h(x,p,t)|$  (here we used that $\Omega$ and $\tilde \Omega$ have the same order of magnitude).
This condition  is satisfied in the vicinity of the extrema of $\mathcal H_{\rm eff}(x,p)$, as long as these  occur in regions of phase space where $\eta |\vec h(x,p,t)|\gg \Omega$. 
As an illustration, in Fig.~\ref{fig:contours}a, we plot the constant (quasi)energy contours of $\mathcal H_{\rm eff}(x,p)$ in the adiabatic regime, obtained from direct numerical evaluation of Eqs.~\eqref{eq:heff}~and~\eqref{eq:heff_expression_adiabatic}. 
We use the      parameters  $A_d= 15$, $B_0=7$, $\omega_c =0.34 \Omega$  and $\eta =0.56 \Omega $.
These parameters are indicated by the cross in Fig.~\ref{fig:Fig1}c, and fall within the adiabatic regime.
%From the considerations above, 
We expect $\mathcal H_{\rm eff}(x,p)$ to describe the system accurately near the three minima located at radius $A_{\rm cav}\approx 24$, where its gradient  (and hence $\delta \omega'$) vanishes.
In Sec.~\ref{sec:numerics} (see also Fig.~\ref{fig:Fig1}c),  we confirm that these local minima indeed lead to quantum frequency locking at these parameters, as explained above.

\subsubsection{Floquet regime}
\label{sec:floquet}
The Floquet parameter regime for frequency locking occurs not when the instantaneous Hamiltonian changes adiabatically, but rather when the  effective {\it Floquet}  Hamiltonian of the spin (with $x$ and $p$ held fixed) changes adiabatically. 

To obtain the effective Hamiltonian $\mathcal H_{\rm eff}(x,p)$ in the Floquet regime, we consider the dynamics resulting from  Eq.~\eqref{eq:s_eom} with  $x$ and $p$ held fixed. 
In this case,  the time-periodicity of $\vec h(x,p,t)$ implies that all solutions $\vec S(t)$ to   Eq.~\eqref{eq:s_eom} satisfy $\vec S([n+1]\tilde T) = R_0(x,p) \vec S(n\tilde T)$,  for some fixed three-dimensional orthogonal matrix $R_0(x,p)$ with unit determinant.
Like any orthogonal matrix with unit determinant, $R_0(x,p)$ can be expressed as a rotation about some axis $\vec a(x,p)$ by some angle $\theta(x, p)$ between $0$ and $\pi$ (note that the required interval for $\theta$  fixes the direction of $\vec a$). % to be in the interval 
As a result, for fixed $x$ and $p$, there exists a   time-periodic solution to the Bloch equation in Eq.~\eqref{eq:s_eom}  (up to a constant scale factor), $\vec{S}(t) = \vec n_0(x,p,t)$, in which $\vec n_0(x,p,0)=\vec a (x,p)$ is parallel to the net rotation axis, and $\vec n_0(x,p,t)$ evolves according to Eq.~\eqref{eq:s_eom}.
Thus, for fixed $x$ and $p$, we  identify $\mathcal H_{\rm eff}^{{\rm spin}} (x,p)= \theta (x,p) \vec a(x,p) \cdot \vec S$ as the effective Hamiltonian of the spin (see Appendix~\ref{app:floquet_locking} for further details). 

When $x$ and $p$ are not fixed, but the evolution of effective precession axis $\vec a(x,p)$ (due to the motion of $x$ and $p$)  evolves slowly relative to the energy gap % effective Larmor precession frequency 
of  $H_{\rm eff}^{{\rm spin}}$, $\delta \varepsilon (x,p) \equiv \min(\theta (x,p),2\pi-2\theta(x,p))/\tilde T$ (see Appendix~\ref{app:floquet_locking}), the stroboscopic motion of the spin closely follows stroboscopic motion resulting from the adiabatically changing Hamiltonian $H_{\rm eff}^{{\rm spin}}(x(t),p(t))$~\cite{Weinberg_2017}. 
As a result, if initially aligned or anti-aligned with $\vec a(x(0),p(0))$, the spin's evolution at later (stroboscopic) times will satisfy $\vec S(n\tilde T) \approx \pm \vec a(x(n\tilde T),p(n\tilde T))$, where the sign depends on the initial alignment. 
%At intermediate times, $\vec S(t) \approx \pm \vec n_0(x( t),p(t),t)
In Appendix~\ref{app:floquet_locking}, we substitute this solution into Eq.~\eqref{eq:classical_rotating_frame_hamiltonian} and take the time-average, making use of our assumption that the cavity mode is effectively stationary within the driving period $\tilde T$~\cite{Oteo_1991}.
Doing this, we find that  the cavity mode evolves according to the  effective Hamiltonian  in Eq.~\eqref{eq:heff}, with 
\be 
\varepsilon (x,p) = \pm  \theta (x,p)/2\tilde T.
\label{eq:heff_expression_floquet}
\ee 
Here the sign depends on the initial alignment of the spin with %the axis
$\vec a(x,p)$~\cite{spin_initialization}. 
The angle $\theta(x,p)$ can be  straightforwardly calculated for the system by exact time-evolution of the Bloch equation for the spin in Eq.~\eqref{eq:s_eom}.

The Floquet regime arises  when the dynamics of the cavity mode occur %take place
on a much longer time-scale  than the driving period $\tilde T$, and when the change of the effective axis of rotation, $\vec a(x,p)$ (due to the motion of $x$ and $p$) is slow compared to the effective Larmor frequency $\theta(x,p)/\tilde T$. 
In Appendix~\ref{app:floquet_locking}, we show that these conditions are satisfied when  $\eta \ll \theta(x,p)/\tilde T$ and $\delta\omega \ll \tilde \Omega/A_{\rm cav}$, where $A_{\rm cav} =\sqrt{x^2+p^2}$ denotes the amplitude of the cavity field. 
Note that, since $\tilde T > T$, and $\theta(x,p)\leq \pi$, the Floquet regime requires $\eta \ll \Omega$. 
Thus the Floquet regime arises in the limit of small spin-cavity coupling, $\eta$, and detuning, $\delta \omega$ (since our semiclassical approximation  requires $A_{\rm cav} \gg 1$ for quantum fluctuations not to play a role).

As an illustration, In Fig.~\ref{fig:contours}b, we plot the contours of $\mathcal H_{\rm eff}(x,p)$ for antialigned spin  (i.e., with $\vec S = -\vec a$) using the parameters $A_d =15, B_0 = 7, \omega_c =\Omega/3$  and $\eta =0.048\Omega $.
Since $\delta \omega =0$,  the conditions for Floquet locking outlined above imply that $\mathcal H_{\rm eff}$  accurately describes the dynamics  of the cavity mode whenever $ H_{\rm eff}(x,p) \gg \eta$.

\section{Quasienergy locking and  symmetry breaking} %Frequency-locked Floquet eigenstates}
\label{sec:phase_locking}
\label{sec:quasienergy_locking}
In Sec.~\ref{sec:theory}, we   identified the conditions for quantum frequency locking: namely, it arises for parameters in the adiabatic or Floquet regimes where 
 the effective   Hamiltonian %of the cavity mode,  
 $\mathcal H_{\rm eff}(x,p)$ [Eq.~\eqref{eq:heff}] has extrema at  nonzero amplitude in phase space.
While our  treatment in Sec.~\ref{sec:theory} in principle also  applies to  fully classical systems (indeed, our results were derived using semiclassical arguments), in this section we demonstrate some novel aspects of the quantum mechanical version
 of the phenomenon. In particular, we show that
 %  interesting on their own}: % that makes it interesting on its own: 
period-$q$ quantum  systems feature 
% reflected in
  characteristic multiplets of  quasienergy levels that differ by rational fractions of the drive frequency, $\Omega/q$, up to exponentially suppressed corrections [see Eq.~\eqref{eq:qel} below]. % for details). 
We term this phenomenon ``quasienergy locking''. 
It was also identified in earlier work~\cite{Holthaus_1994,Sacha_2015,Else_2016b,Zhang_2017}, and
can be seen as  the defining feature of quantum frequency locking.
%As also discussed in Refs.~\cite{Sacha_2015,Zhang_2017}, q

Quasienergy locking can be seen as a breakdown of the discrete time translation symmetry which is  inherently present in the %periodically 
driven 
system~\cite{Sacha_2015,Else_2016b,Zhang_2017}: 
%specifically,
 as we show below, 
linear combinations of  quasienergy-locked Floquet eigenstates define a family of nearly stationary states of the system's  evolution that break the discrete time translation symmetry of the drive (up to exponentially long times). %; this feature is %equivalently

In the remainder of this section we  review the defining features of quasienergy locking (Sec.~\ref{sec:qel_def}), and subsequently show how it arises the qubit-cavity model (Sec.~\ref{sec:qel_derivation}).
We finally discuss % in further detail
 how quasienergy locking can be understood as a breakdown of discrete time translation symmetry (Sec.~\ref{sec:ttsb}).
% (see also, e.g. Refs.~\cite{Holthaus_1994,Sacha_2015,,Zhang_2017})
To highlight physical aspects of the phenomenon, we provide our arguments on a heuristic level, while  a   more rigorous (but  technical) treatment  is given in Appendix.~\ref{app:qe_locking}.
%translation symmetry of the system~\cite{Sacha_2015,Zhang_2017}.
%

\subsection{Quasienergy locking}
\label{sec:qel_def}
%Here we review the defining features of quasienergy locking .
%As explained above, q
Quasienergy locking is a phenomenon that arises in  the quasienergy spectrum of % the  Floquet eigenstates   and quasienergies in
 periodically driven quantum systems~\cite{Holthaus_1994,Sacha_2015,Else_2016b,Zhang_2017}. 
% (such as the qubit-cavity system we consider).
%For this class of 
In such systems the quasienergies $\{\varepsilon _n\}$ define % the Floquet eigenstates $\{|\psi_n\rangle\}$ form a complete basis of states that are mapped to themselves after each driving period $T$, up to a unitary phase that defines the corresponding quasienergies  $\{\varepsilon _n\}$: %(see Sec.~\ref{sec:floquet_intro} for their definition): 
the eigenvalues of the system's time-evolution operator over one period (known as the Floquet  operator),  $\hat U(T) |\psi_n \rangle = e^{-i\varepsilon _n T}|\psi_n\rangle$, where $\hat U(t) \equiv Te^{-i\int_0^t dt'\hat  H(t')}$ denotes the system's time-evolution operator and $\mathcal T$ denotes the time-ordering operation. 
%Specifically,. 
The corresponding  eigenstates $\{|\psi_n\rangle\}$, termed Floquet eigenstates, hence form a complete basis of states that are mapped to themselves after each driving period $T$, up to a unitary phase $e^{-i\varepsilon _nT}$. % defined by  $\varepsilon _n$.
Quasienergy thus plays a role analogous to energy for the evolution of periodically driven quantum systems: 
%Analogous to non-driven systems, 
the evolution at  integer  multiples of the driving period, $k$, can  be resolved   as %terms of the Floquet eigenstates and quasienergies:
$
|\psi(kT)\rangle = \sum_n c_n e^{-i\varepsilon _n kT}|\psi_n\rangle,
$
where %, %as for nondriven systems, 
the coefficients 
$\{c_n\}$ are determined from the initial conditions~\cite{floquet_propagation}. 
However, note that each $\varepsilon _n$ is only defined modulo $\Omega$. % $c_n = \langle \psi_n|\psi(0)\rangle$. 
%In this way, quasienergy plays a role analogous to energy for the evolution of the system; however, it is only defined modulo $\Omega$.

%\addIM{IM: this does not define chi's fully, since the phase of psi is free. What we want is similar to the construction of the maximally localized Wannier functions}. 
% \FNcomment{Blue part below is modified  in response. Ok?}
%While for generic periodically driven systems,
The quasienergies of a generic periodically driven quantum system %$\{\varepsilon _n\}$
 are naturally distributed uniformly  between $0$ and $\Omega$. % in the interval $[0,\Omega]$. 
However, 
%in Sec.~\ref{sec:qel_derivation} below, we show how 
when period-$q$ quantum frequency locking  arises, the spectrum features %results in the formation of
 characteristic multiplets of  %Floquet eigenstates  whose  quasienergies 
quasienergy levels $\varepsilon _1 ,\ldots \varepsilon _q$ that  differ by  $\Omega/q$, %a rational fraction  of the driving frequency,
  up to exponentially suppressed corrections: 
\be 
\varepsilon _\ell = \varepsilon  + \ell \Omega/q +\mathcal O(\delta \varepsilon ),
\label{eq:qel}
\ee
% (See also introduction and Fig.~\ref{fig:Fig1}b.)
where $\varepsilon$ generally differs from multiplet to multiplet.
Here $\delta \varepsilon $ is a quasienergy scale that can be many times %orders of magnitude
 smaller than the  average quasienergy level spacing in the system, such that the feature above would not occur by coincidence  (see, e.g., the inset in Fig.~\ref{fig:Fig1}c and Sec.~\ref{sec:numerics}). %,
For the driven qubit-cavity system we consider in this work, we show below that
%As we show below, for the driven qubit-cavity system we consider,
 $\delta \varepsilon \sim e^{-d/\xi}$, where $d$ denotes the separation  in phase space between the extrema of the effective Hamiltonian 
 $\mathcal H_{\rm eff}(x,p)$, while $\xi\sim 1$ denotes the scale of quantum fluctuations. 
We term the  formation of these multiplets ``quasienergy locking''.

The  Floquet eigenstates associated with  each multiplet of  locked quasienergy levels,  $|\psi_1\rangle,\ldots |\psi_q\rangle$, have interesting features of  their own.
Specifically, they take the form
%we may write them in the form
\be 
|\psi_\ell\rangle = \frac{1}{\sqrt{q}}\sum_{k=1}^q e^{-2\pi i \ell k /q}|\chi_k\rangle,
\label{eq:qel_fes_form}
\ee
where, for the model %driven qubit-cavity system
we consider, %we find below that
 each state $|\chi_k\rangle$  has support only near a particular extremum of $\mathcal H_{\rm eff}(x,p)$ in phase space  (see Appendix~\ref{app:qe_locking} and Sec.~\ref{sec:qel_derivation} below).
Note from Eq.~\eqref{eq:qel_fes_form}  that, for each $k$, $|\chi_k\rangle \equiv \frac{1}{\sqrt{q}}\sum_{k} e^{2\pi i \ell k /q}|\psi_k\rangle$.
One can thus verify that  the states $|\chi_1\rangle \ldots |\chi_k\rangle$  are  orthogonal, and mapped to each other under evolution by one driving period $T$, up to a phase and an exponentially suppressed correction:
\be 
\hat U(T) |\chi_k\rangle = e^{-i\varepsilon T} |\chi_{k+1}\rangle  + \mathcal O (\delta \varepsilon  T),
\label{eq:chi_relation}
\ee
with $|\chi_{q+1}\rangle \equiv |\chi_q\rangle$~\cite{other_families}.

\subsection{Derivation of quasienergy locking for the driven qubit-cavity system} %Frequency-locked Floquet eigenstates}
\label{sec:qel_derivation}
Having reviewed the defining features of quasienergy locking, we now   show how it  emerges in the driven qubit-cavity system that we consider in this work.
%We provide our arguments on a heuristic basis, while  a more more rigorous derivation is given in  Appendix~\ref{app:qe_locking}. % for a more more rigorous discussion).
%To show how the quasienergy locking emerges, % phenomenon emerges in the driven qubit-cavity model,
We provide our arguments on a heuristic basis, by analyzing the dynamics of the system in the rotating frame introduced in Sec.~\ref{sec:effective_hamiltonian}. 
For simplicity we consider  the   limit %  \addMR{\bf [MR: specify limit?]}
 where the %dynamics 
{evolution} of the system in this %rotating 
frame % (see Sec.~\ref{sec:effective_hamiltonian} 
is  fully captured by the effective semiclassical  Hamiltonian from Sec.~\ref{sec:theory}, $\mathcal H_{\rm eff}(x,p)$ (see Sec.~\ref{sec:effective_hamiltonian} for specific conditions). %\comment{ok?}
In Appendix~\ref{app:floquet_locking} we provide a more  rigorous line of arguments that 
% also hold beyond this limit. 
%Specifically, the derivation in Appendix~\ref{app:floquet_locking}
% applies to all cases where 
 holds in the Floquet and adiabatic limits whenever $\mathcal H_{\rm eff}(x,p)$ features extrema with surrounding ``potential wells''  that are much larger than the scale of quantum fluctuations, $\xi \sim 1$ (see below for definition of potential wells).
% the  dynamics generated by the Hamiltonian in   the rotating frame, $\tilde H(t)$ [see Eq.~\eqref{eq:rotating_frame_hamiltonian}], features  solutions at finite displacement amplitude, $ \sqrt{x^2+p^2}$, that remain near-stationary in phase space up to exponentially long timescales. % that remain nearly stationary. 

%Our goal is to obtain 
%As explained in Sec.~\ref{sec:qel_def}, quasienergy locking is a feature of the quasienergy spectrum and Floquet eigenstates of the system (in the lab frame).
%To establish the emergence of quasienergy locking in the driven qubit-cavity system,
%As a first step
To derive quasienergy locking %  we  obtain the quasienergy spectrum and Floquet eigenstates of the system
%from considerations about
%we study
we investigate  the properties of the driven qubit-cavity system's Floquet eigenstates and quasienergies, $\{| \psi_n\rangle\}$ and $\{\varepsilon _n\}$.
%and $\{\tilde \varepsilon _n\}$.
As a first step, we note that the former are identical to the Floquet eigenstates in the rotating frame (see Sec.~\ref{sec:effective_hamiltonian}), $\{|\tilde \psi_n\rangle\}$, %are identical to those in the rotating frame, 
while each quasienergy % in the lab frame,
 $\varepsilon _n$ is identical to  the corresponding quasienergy  in the rotating frame, $\tilde \varepsilon _n$, modulo $\tilde \Omega \equiv \Omega / q$~\cite{qe_correspondence}.
%% $\varepsilon _n = \tilde \varepsilon _n (\mod \tilde \Omega)$.
%This follows since the Floquet operator in the rotating frame, ${ {\tilde U}}(qT)$, is given by %$, is identical to $q$th power of the Floquet operator in the lab frame, $
%$[\hat U(T)]^q$, where $\hat U(T)$ denotes the Floquet operator in the lab frame (recall that $ {\hat U}(t)=\hat U_0(t) \tilde U(t)$, while $\hat U_0(qT) = 1$). 
%In the limit where  
To obtain %he Floquet eigenstates and quasienergies in the rotating frame,  
$\{|\tilde \psi_n\rangle\}$ and $\{\tilde\varepsilon _n\}$, we 
we recall  our assumption that  $\mathcal H_{\rm eff}(x,p)$ fully captures the system's dynamics in the rotating frame. 
%We thus expect 
%Hence, 
Hence we expect the stroboscopic evolution of the quantized cavity mode (in the rotating frame) to be generated by the   effective quantum  Hamiltonian $\mathcal H_{\rm eff}(\hat x,\hat p)$.
We moreover expect the spin to be locked to the effective axis of precession as a function of $ x$ and $ p$, as explained in Sec.~\ref{sec:theory}.
For simplicity, we therefore  neglect the spin in the following. 
Through the above correspondence between the  rotating and lab frames, in the idealized limit we consider here, the Floquet eigenstates of the system in the lab frame are thus
given by  the eigenstates of $\mathcal H_{\rm eff}(\hat x,\hat p)$, while the quasienergies are given by the corresponding eigenvalues, up to integer multiples of $\tilde \Omega$.

%As explained in
To obtain the eigenstates and eigenvalues of  $\mathcal H_{\rm eff}(\hat x,\hat p)$, %n the  regime,  we 
 we consider the structure of $\mathcal H_{\rm eff}(\hat x,\hat p)$ in the frequency-locked regime.
 We  recall %from  Sec.~\ref{sec:theory} 
%that frequency locking occurs %in the qubit-cavity system
% for parameters where
% we recall 
 that frequency locking arises when $\mathcal H_{\rm eff}(x,p)$ has extrema at nonzero displacement amplitude. %$\sqrt{x^2+p^2}$.
These extrema are surrounded by {classical} trajectories [i.e., contours of $\mathcal H_{\rm eff}(x,p)$] %constant effective energy contours)}
 which encircle and remain close to their respective fixed points at all times.
We refer to each such extremum, along  with its surrounding neighborhood that contains these encircling trajectories (out to the separatrices beyond which the trajectories encircle other fixed points) as a potential well.
Note that  $\mathcal H_{\rm eff}(x,p)$ has a built-in symmetry of discrete rotation  by   $2\pi/q$ in phase space, as is evident in the numerical examples plotted in Fig.~\ref{fig:contours}ab, where $q=3$ (see also Appendix~\ref{app:qe_locking}). 
This symmetry, which is generated by $\hat U_0(T)$, %and corresponds to discrete time-translation symmetry by $T$, 
guarantees that each potential well of $\mathcal H_{\rm eff}(x,p)$  forms part of a ring of $q$  wells which are mapped to each other through rotations by  $2\pi/q$ in phase space.
We refer to these  wells as wells $1\ldots q$ in the following, such that well $k$ is mapped to well $k+1\, (\mod q)$ a  through phase space rotation by $2\pi /q$.

% \addIM{IM: would be good to explain the difference between phi and chi wavefunctions: on in rotating frame, the other in Lab?} 
% \FNcomment{Text modified below; ok?}
% We now use  the above  properties of $\mathcal H_{\rm eff}(\hat x,\hat p)$  to infer the structure of its eigenstates and eigenvalues  in the frequency-locked regime (and hence also of the Floquet eigenstates and quasienergies in the lab frame). %& and hence also of  the  of the qubit-cavity system. 
% in the lab frame.
% In the % regime, % 
%  regime (where potential $\mathcal H_{\rm eff}(x,p)$ features potential wells), 
We  expect % the quantized effective Hamiltonian 
 $\mathcal H_{\rm eff}(\hat x, \hat p)$   to support %``bound'' 
{approximate} eigenstates   that are confined within the potential wells of $\mathcal H_{\rm eff}(x,p)$, whose wave functions  decay with the distance from the well, $r$, as $\mathcal O(e^{-r/\xi})$. % where $r$ denotes the distance to the well in phase space.
For $k=1,\ldots q$, we let $|\phi_k\rangle$ % an even-amplitude linear combination of  $q$  bound states $|\phi_1\rangle,|\ldots \phi_q\rangle$, where $|\phi_k\rangle$  
  denote such a ``bound'' eigenstate of $\Heff(\hat x , \hat p)$ when  restricting phase space to well $k$, such that $|\phi_k\rangle$ and $|\phi_{k+1}\rangle$ are related through phase space rotation by $2\pi /q$; i.e., we restrict phase space to a  region located within a distance $r_0$ from well $k$, for some $0<r_0<d/2$, where $d$ denotes the distance  in phase space between adjacent wells.
  We let $\varepsilon $ denote the corresponding eigenvalue of $H_{\rm eff}(\hat x ,\hat p)$; this takes  the same value for all $k$ due to the discrete rotation symmetry.
Below we find that the bound states $\{|\phi_k\rangle\}$ are identical to the % |\chi_k\rangle$, where % the states 
%$\{|\chi_k\rangle\}$ are the 
states  $\{|\chi_k\rangle\}$ whose linear combinations give %which together span %form the  separate (in phase space) components of 
a quasienergy-locked multiplet of  Floquet eigenstates  as in Eq.~\eqref{eq:qel_fes_form}. 
Note, however, that this identification is only exact in the idealized limit we consider here, where  $ {\mathcal H}_{\rm eff}(\hat x, \hat p)$ fully captures the stroboscopic evolution in the rotating frame.
Beyond this limit,  $\{|\chi_k\rangle\}$ %from Sec.~\ref{sec:qel_def} 
deviate from  $\{|\phi_k\rangle\}$ by  nonzero, but small, corrections due to, e.g., nonadiabatic corrections and quantum fluctuations.
In Appendix~\ref{app:qe_locking}, we provide a more rigorous way to identify the states $|\chi_n\rangle$ that includes such corrections.

  Due to the exponentially decaying wavefunction of $|\phi_k\rangle$ outside well $k$, when all of phase space is included, each $|\phi_k\rangle$
  remains an approximate eigenstate of $\Heff(\hat x, \hat p )$,  up to an exponentially small correction
%\addMR{\bf [MR: if true eigenstates are equal weight superpositions, how is this consistent?]}: \FNcomment{Ok?}
 $\Heff(\hat x, \hat p )|\phi_k\rangle = \varepsilon |\phi_k\rangle + \mathcal O(e^{-r_0/\xi})$. 
 To zeroth order in  $\lambda  =e^{-r_0/\xi}$ (i.e., in the classical limit $\xi \to 0$), each $|\phi_k\rangle$ is thus an exact eigenstate of $\Heff(\hat x , \hat p)$, with eigenvalue $\varepsilon $. 
 In the limit of small but nonzero $\xi$, the corresponding eigenstates of $\Heff(\hat x ,\hat p)$, $|\psi_1\rangle,\ldots|\psi_q\rangle$,
 %are linear combinations of the degenerate ``unperturbed'' states $|\phi_k\rangle$ a
 can be obtained through
  zeroth-order degenerate perturbation theory, and thus can be expressed as linear combinations of the degenerate ``unperturbed'' states $|\phi_k\rangle$. 
%  
% The coefficients of these linear combinations  can be obtained through zeroth order perturbation theory:
To identify these linear combinations, we note that the discrete rotation  symmetry  $[\hat U_0(T),\mathcal H_{\rm eff}(\hat x,\hat p)]=0$  requires the  eigenstates of $\mathcal H_{\rm eff}(\hat x,\hat p)$ to also be eigenstates of $\hat U_0(T)$.
Since $\hat U_0(T) |\phi_k\rangle  = |\phi_{k+1}\rangle$, %  the  eigenstates of $\mathcal H_{\rm eff}(x,p)$, 
$|\psi_1\rangle,\ldots |\psi_q\rangle$ are thus given by Eq.~\eqref{eq:qel_fes_form}, with $|\chi_n \rangle = |\phi_n\rangle$. 
Using that $\langle \phi_k|\Heff(\hat x, \hat p)|\phi_{k'}\rangle \lesssim e^{-d/\xi}$, we find that the corresponding eigenvalues of $\Heff(\hat x ,\hat p)$,  $\tilde \varepsilon _1,\ldots \tilde \varepsilon _q$, are all given by $\varepsilon $, up to corrections of order $e^{-d/\xi}$. 
We conclude that for
 each ring of potential wells  of $\Heff(x,p)$ (if such wells are present), %,
  the qubit-cavity system %in the lab frame 
  supports one or more families of  Floquet eigenstates, whose form is given in Eq.~\eqref{eq:qel_fes_form}, where, for each $k$, $|\chi_k\rangle$ has support only in well $k$ of the ring~\cite{more_families} . 
The corresponding  quasienergies are given by the same value $\varepsilon $, up to integer multiples of $\Omega /q$, and corrections of order $e^{-d/\xi}$.

In Appendix~\ref{app:qe_locking}, we provide a  more rigorous % \addMR{rigorous? let's not imply that we're not careful...}
 derivation of quasienergy locking, which holds beyond the idealized limit considered here. 
 There we confirm that in the  regime, the qubit-cavity system supports families of near-degenerate eigenstates of the form in Eq.~\eqref{eq:qel_fes_form}, where $|\chi_k\rangle$ has support only in well $k$. 
However,   $|\chi_k\rangle$ and $|\chi_{k+1}\rangle$ are not  related by  exact phase space rotation by $2\pi/q$, but rather through Eq.~\eqref{eq:chi_relation}~\cite{approximate_rotation}.
It follows that the corresponding quasienergies take the form in Eq.~\eqref{eq:qel}. 
This was what we wanted to show in this subsection.

%hence   Eq.~\eqref{eq:chi_relation} is fully consistent with $|\chi_k\rangle$ having support only in well $k$ for each $k$. 
%, and hence by construction is a symmetry  of $ the exact effective Hamiltonian in the rotating frame. 

%  the  state's location in phase space by much.

%The above derivation shows  that, when present, each ring of potential wells of $\Heff(x,p)$ support one or more families of  Floquet eigenstates of the lab frame system, whose form is given in Eq.~\eqref{eq:qel_fes_form}, while the corresponding quasienergies are are given by Eq.~\eqref{eq:...}.
%To determine the integer $z_\ell$, we use that $\hat U (T) |\chi_k\rangle =  e^{-\varepsilon T}|\chi_{k+1}\rangle$.  
% 
\subsection{Time translation symmetry breaking}
\label{sec:ttsb}\label{sec:qel_signatures}
Here we review how quasienergy locking can be seen as a realization of time-translation symmetry breaking~\cite{Sacha_2015,Else_2016b,Zhang_2017}.
To see this,  note from Eq.~\eqref{eq:chi_relation} that the states $\{|\chi_k\rangle\}$ are taken onto themselves after evolution by the extended period $\tilde T = qT$, up to a phase, and an exponentially suppressed correction.
Each $|\chi_{k}\rangle$ hence is a  (nearly) stationary state of the system's time-evolution that breaks the original discrete time-translation symmetry by $T$.
In contrast to the exact Floquet eigenstates, $\{|\psi_\ell\rangle\}$, which are
 superpositions of states characterized by distinct values of the oscillator phase (i.e, ``{Schr\"{o}dinger} cat'' states), each symmetry-breaking state $|\chi_k\rangle$  has a well-defined phase, and hence corresponds to a semiclassical ``non-cat'' state of the cavity mode.
In the sense above, the driven qubit-cavity system can hence be seen as supporting  steady states  that break  discrete time translation symmetry.

The symmetry-breaking states $\{|\chi_k\rangle\}$  remain stationary states of the system's time-evolution up to the duration of  confinement within the potential wells of $\mathcal H_{\rm eff}$,  $\tau~\sim e^{{d/\xi}}$. 
In contrast, the ``coherence time'' of the symmetry-breaking steady states in many-body Floquet time crystals scales exponentially with the  size of the system due to many-body nature of the states, and hence is infinite in the thermodynamic limit~\cite{Khemani_2016,Else_2016b}. 
For the qubit-cavity system we consider, {although there is no thermodynamic limit}, $\tau$ still scales exponentially with the system parameters, and thus can be very large compared to the other timescales of the system. 
In this sense, we can still regard time-translation symmetry to be  broken in practice.
Note that we consider a more general notion of time-translation symmetry breaking than defined in Ref~\cite{Else_2016b}: namely, we only require some, but not  all,  steady states of the system to break the discrete time-translation symmetry of the system~\cite{Sacha_2015,Zhang_2017}.

\section{Numerical results\label{sec:numerics}}

Here we support our %the
discussion %in Sec.~\ref{sec:theory}  by
with numerical simulations. 
%of the qubit-cavity system we consider in this paper (see Sec.~\ref{sec:model}). 
We simulate the qubit-cavity model in Sec.~\ref{sec:model} by computing the complete Floquet operator of the system % (see Sec.~\ref{sec:model}) 
using direct time-evolution.
We then obtain the quasienergy spectrum and Floquet eigenstates through exact diagonalization. %\ref{eq:j_sc_def}). 
%The only approximations we make  in our 
In these simulations we  truncate  the Hilbert space of the cavity to the first $650$ photon-number eigenstates  (resulting in  Hilbert space  dimension $1300$), and discretize the  Hamiltonian's continuous time-dependence within one period into  $300$ evenly spaced intervals.

Throughout our simulations, we fix the dimensionless Zeeman field component to be $B_0=7$, and the driving amplitude  $A_d=15$, while we vary the qubit-cavity coupling $\eta$ and the cavity resonance frequency 
$\omega_{\rm c}$ (see Sec.~\ref{sec:model}).

\subsection{Detection of quantum frequency locking}
\label{sec:locking_detection}
For each choice of the parameters $\eta$ and $\omega_{\rm c}$ we probed, we detected the presence of quantum frequency locking from the quasienergy spectrum of the system, $\{\varepsilon _n\}$.
We begin by sorting the quasienergy level spacings for the system, 
$\Delta \varepsilon _{mn} \equiv \varepsilon _{m}-\varepsilon _{n}$ for all $1300\times 1299$  pairs of quasienergy levels where $m\neq n$, into a histogram of $10^5$ bins evenly spaced in the interval between $0$ and $\Omega$ (we consider  the value of each level spacing $\Delta \varepsilon _{mn}$  modulo $\Omega$).
For a generic distribution of quasienergy levels, we  expect the   number of level pairs  $N(\Delta \varepsilon )$  falling into the  bin at level splitting $\Delta \varepsilon $ 
to be given by $ 1300^2 / 10^{5} \approx 17$. 
However, when period-$q$ quantum frequency locking is present, we expect an anomalously high number of level spacings to fall into the bin where $\Delta \varepsilon  = \Omega /q$, c.f.~the discussion in Sec.~\ref{sec:phase_locking}. 

To illustrate this, in the inset in Fig.~\ref{fig:Fig1}c, we plot $N(\Delta \varepsilon )$ for %the parameters
$\eta  = 0.56 \Omega$ and $\omega_{\rm c} = 0.34 \Omega$ (indicated by cross in main panel).
These parameters bring the system into the adiabatic regime; we previously plotted the $q=3$ effective cavity Hamiltonian for this choice of parameters in Fig.~\ref{fig:contours}a (see Sec.~\ref{sec:effective_hamiltonian}). 
From the arguments of Sec.~\ref{sec:theory}, the  local extrema of $\mathcal H_{\rm eff}(x,p)$, which are clearly present in Fig.~\ref{fig:contours}a, should give rise  to period-$3$ frequency locking. 
The data in the inset of  Fig.~\ref{fig:Fig1}c confirms this: 
while $N(\Delta \varepsilon )$ is of order $\sim 17$ for almost all bins in Fig.~\ref{fig:Fig1}c, the spectrum features an anomalously high number of level pairs ($\sim 100$) %in the spectrum
whose splitting falls into the bin at $\Delta \varepsilon =\Omega /3$. 
From the discussion in Sec.~\ref{sec:phase_locking}, this  is a clear indication of period-$3$ quantum frequency locking. % in the model for these parameters. 
We expect the model supports  approximately $100-17 \sim 85$ frequency-locked triplets of  Floquet eigenstates of the form in Eq.~\eqref{eq:triplet}.

As the above paragraph demonstrates, we may use the histogram peak-height
$N(\Delta \varepsilon =\Omega/q)$  to estimate the number of period-$q$ frequency-locked Floquet eigenstates  in the system. 
In the main panel of Fig.~\ref{fig:Fig1}c, we plot this  number for $q=3$ as a function of $\omega_{\rm c}$ and  $\eta$. 
As is evident in Fig.~\ref{fig:Fig1}c, the model supports a large number of period-$3$ frequency-locked Floquet eigenstates in a finite region of parameter space, arising both for weak and strong detuning $\delta \omega = {\omega_{\rm c}}-\Omega/3$ and qubit-cavity coupling $\eta$.

The data in Fig.~\ref{fig:Fig1}c show clear signatures of the two distinct regimes of quantum frequency locking we identified in Sec.~\ref{sec:effective_hamiltonian}. 
Focusing on the peak  that emerges from $\omega_{\rm c} = \Omega/3$, for  $\eta\ll \Omega$,  quantum frequency locking occurs when $\omega_{\rm c}\approx \Omega /3$.
However,  for $\eta\gtrsim\Omega/2$, the $\omega_{\rm c}$-interval in which quantum frequency locking occurs splits into two linearly-diverging branches.
This  point   marks the crossover from the Floquet  (lower branch) to the adiabatic regime (upper branches).
Specifically, in the adiabatic regime, after  a simultaneous  rescaling of $\eta$ and $\delta \omega$ by the same positive factor $\lambda$, $\mathcal H_{\rm eff}(x,p)$ is mapped to $\lambda\mathcal H_{\rm eff}(x,p)$ [see Eqs.~\eqref{eq:heff}~and~\eqref{eq:classical_rotating_frame_hamiltonian}].
Moreover a sign reversal of $\delta \omega$ maps $\mathcal H_{\rm eff}(x,p)$ for aligned spin into $-\mathcal H_{\rm eff}(x,p)$ for anti-aligned spin, and vice versa. 
Thus, in the adiabatic  regime, $\mathcal H_{\rm eff}(x,p)$  features the same structure of local extrema and  potential wells along the lines $ \delta \omega = \pm \kappa \eta$ for some proportionality factor $\kappa$, and hence,  in this regime, quantum frequency locking should occur along these two lines in parameter space.
This structure of  linearly-diverging branches is clearly evident in Fig.~\ref{fig:Fig1}c. 
In contrast, the Floquet  regime only arises when $\eta < \Omega $, and for small values of $\delta \omega$ (see Sec.~\ref{sec:floquet}).
Thus, %
%while the adiabatic regime gives rise to two branches of frequency locking in parameter space at detuning $\delta \omega \propto \pm \eta$, 
the Floquet  regime gives rise to a single branch  at
$\delta \omega \sim 0, \eta \ll \Omega$.

\subsection{Structure of Floquet eigenstates}
% As a next goal, %for our simulations,
Next, we sought to verify that the frequency-locked Floquet eigenstates have the structure we predicted in Sec.~\ref{sec:phase_locking}: we expect each triplet of  frequency-locked Floquet eigenstates,  $|\psi_n^1\rangle$, $|\psi_n^2\rangle$, and $|\psi^3_n\rangle$ (with corresponding quasienergies $\varepsilon_n + \ell \Omega /3$) to be of   the form in Eq.~\eqref{eq:floquet_state_multiplet_final} (for $q=3$),
% \be 
% |\psi_n^\ell\rangle  =\frac{1}{\sqrt{3}}(\alpha^{\ell}|\chi^1_n\rangle + \alpha^{2\ell}  |\chi^2_n\rangle + \alpha ^{3\ell} |\chi^3_n\rangle),
% \label{eq:hypothesized_structure}
% \ee
%where $\alpha = e^{-2\pi i /3}$, and 
where $|\chi_n^k\rangle$  has support only within a particular ``potential well'' of the effective cavity mode Hamiltonian, $\mathcal H_{\rm eff}(x,p)$. 

To confirm %that the frequency-locked Floquet states exhibit 
the hypothesized structure above, we  obtained the Floquet eigenstates of the model for the  parameter set  used  in  Fig.~\ref{fig:contours}a (see Sec.~\ref{sec:theory} for parameters; note that these were also used for the inset in Fig.~\ref{fig:Fig1}c), where the system exhibits quantum frequency locking in  the adiabatic regime. 
We computed the Wigner function $W(x,p)$ for each Floquet eigenstate $|\psi_n\rangle$,
using  the reduced density matrix  of the cavity  $\rho^{n}_{\rm cav}= \Tr_S [|\psi_n\rangle\langle\psi_n|]$, where $\Tr_S$ denotes the partial trace over the Hilbert space of the spin. 
Fig.~\ref{fig:contours}c shows the Wigner function of a frequency-locked  Floquet eigenstate of the system (i.e.,  one out of the %set of
many Floquet eigenstates whose  quasienergies differ by an exact multiple of $\Omega/3$ from two other quasienergies in the system). 
The Wigner function  % of the  Floquet eigenstate 
in Fig.~\ref{fig:contours}c shows a highly structured pattern, and has  support only in $3$ separate regions of phase space that coincide with %the locations of 
the potential wells %in phase space 
of  $\mathcal H_{\rm eff}(x,p)$ in panel (a),  (shown as grey lines in Fig.~\ref{fig:contours}c), consistent with Sec.~\ref{sec:phase_locking}. 

Next, we identified the   two other Floquet eigenstates of the triplet in which $|\psi_n\rangle =|\psi_n^3\rangle$ formed a part, $|\psi_n^1\rangle$ and $|\psi_n^2\rangle$ (i.e., we identified the two Floquet eigenstates of the system whose   quasienergies differ  by  $\pm\Omega/3$ from the quasienergy of the eigenstate $|\psi_n^3\rangle$, up to a correction many orders of magnitude smaller than the level spacing of the quasienergy spectrum).
The  Wigner functions of these two  states are nearly identical to the Wigner function  in Fig.~\ref{fig:contours}c, and are not shown here. 
According to the hypothesis of Eq.~\eqref{eq:floquet_state_multiplet_final} there exists a gauge choice for the Floquet eigenstates $|\psi_n^1\rangle,\ldots|\psi_n^3\rangle$  such that, for each $k$, $|\chi^k_n\rangle \equiv \frac{1}{\sqrt{3}}\sum_{\ell=1}^3e^{2\pi i k\ell/3}|\psi_n^\ell\rangle$  only has support only in well $k$ of $\mathcal H_{\rm eff}$. 
% We computed the  equal-weight  linear combinations of the three Floquet eigenstates, $\frac{1}{\sqrt{3}}\sum_{\ell=1}^3e^{2\pi i k\ell/3}|\psi_n^\ell\rangle$  for $k=1,2,3$. 
In the inset of Fig.~\ref{fig:contours}c, we show the Wigner function for such a linear combination  (with $k=3$). 
In agreement with the discussion in Sec.~\ref{sec:phase_locking},  %t
this Wigner function is only nonzero in a single potential well  of $H_{\rm eff}$ [namely near $(x,p)= (20,0)$].
We confirmed numerically (data not shown here) that with the same gauge choice for the states $|\psi_n^1\rangle,\ldots |\psi_n^3\rangle$, % as used in Fig.~\ref{fig:contours}c,
the two other choices of $k$ led to the  Wigner function of the resulting state $|\chi_k\rangle$ having support in the two other potential wells of $H_{\rm eff}$.
Thus we confirmed that the triplet of Floquet eigenstates has the structure in Eq.~\eqref{eq:floquet_state_multiplet_final}.

We also considered the Wigner functions of frequency-locked Floquet eigenstates in the  Floquet regime ($\eta \ll \Omega, \delta \omega \sim 0$). 
Fig.~\ref{fig:contours}d shows the Wigner function of such a frequency-locked Floquet eigenstate of the model, for  parameters  $ \omega_{\rm c} =\Omega/3$ and $\eta =0.048\Omega $, which puts the system in the Floquet regime, and were also used in Fig.~\ref{fig:contours}b.
The  Wigner function %s of the   Floquet state 
exhibits a very similar structure as in the adiabatic regime: there exist three separate regions where the it is nonzero and smoothly varying that coincide with the potential wells  of the effective Hamiltonian of the system,  $\mathcal H_{\rm eff}(x,p)$, (Fig.~\ref{fig:contours}b). 
At the edges of its peaks, the Wigner function   exhibits oscillations from positive to negative with nodal lines parallel to the  the contours of  $\mathcal H_{\rm eff}(x,p)$,  hence strongly supporting the discussion in Sec.~\ref{sec:floquet}. % \ref{sec.}

Note that each  ring of potential wells of $\mathcal H_{\rm eff}(x,p)$ % triplet of the effective cavity  mode Hamiltonian $\mathcal H_{\rm eff}(x,p)$
can support several triplets of quasienergy-locked  Floquet eigenstates.
Moreover, for the (Floquet regime) parameters used in Fig.~\ref{fig:contours}bd,  $\mathcal H_{\rm eff}(x,p)$ features multiple rings of potential wells that can each support its own
%; each of these  %rings in turn supports its own 
families of frequency-locked Floquet eigenstates. 
We confirm this in Appendix~\ref{app:other_wigners} where   we plot the Wigner functions for  two additional triplets of frequency-locked Floquet eigenstates that both have support in the same well; this well  is different from the one where the Floquet eigenstate   in Fig.~\ref{fig:contours}d has support. 
%while the second has support in a different well. 

\subsection{Observable signatures and frequency conversion}\label{sec:signatures}

%%%%%%%%%%%%%%%%%%%%%%%%%%%%%%%%%%%%%%%%%%%%%%%%%%%%%%%%%%%%%%%%%%%%%%
\begin{figure}
\includegraphics[width=0.99\columnwidth]{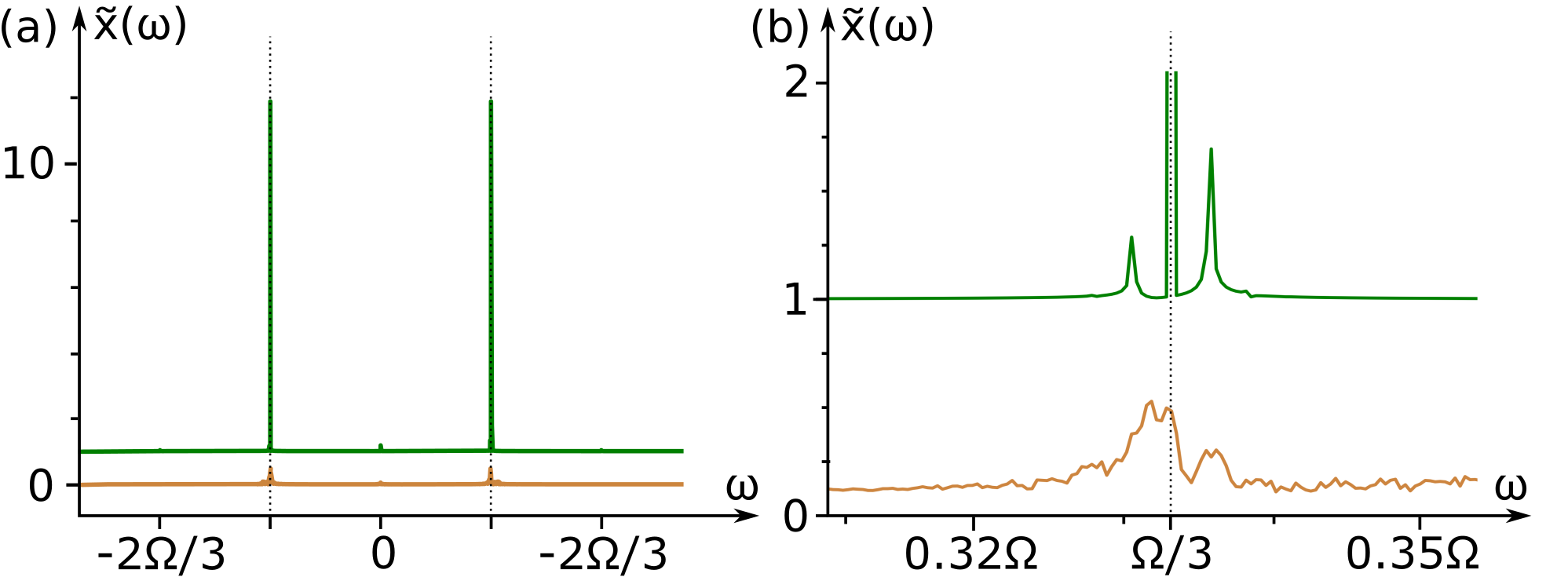}
\caption{
Observable signatures of quantum frequency locking. 
(a) Frequency spectrum of the cavity field, $\langle \hat x(t)\rangle $, for the system when initialized within (green), and outside (orange) the frequency locking regime,  respectively.
See main text for parameters and further details.
Note that  the green curve is vertically offset by $1$.
(b) Zoom-in of panel (a), in the vicinity of  $\omega = \Omega/3$ (indicated by vertical dashed line). }
%\comment{Change to .pdf figure in final version to make lines go away}
%\comment{Zoom in on panel c -- make it go to 2 Omega.}}
\label{fig:dynamics}
\end{figure}
%%%%%%%%%%%%%%%%%%%%%%%%%%%%%%%%%%%%%%%%%%%%%%%%%%%%%%%%%%%%%%%%%%%%%%
As a final goal for our numerical simulations, we explored  the observable signatures of quantum frequency locking in the system, and their possible applications for frequency conversion. 
To this end, we  considered the dynamics of the % cavity mode %physical
observable $\langle \hat x(t)\rangle$, which, depending on the exact realization of the model,  %of this quantity. 
 for instance can measure a component of the electromagnetic field in the cavity  (see Sec.~\ref{sec:model} for definition). 
Using the parameters $\omega = 0.34 \Omega$, $\eta = 0.56 \Omega$ %(%these were
(also used in Figs.~\ref{fig:Fig1}c~and~\ref{fig:contours}ac), 
we computed   the time-evolution of the system after initializing the cavity mode in a coherent state with phase $0$ and displacement amplitude either $20$ or $10$, corresponding to locations $(x_0,p_0)=(20,0)$ and $(x_0,p_0)=(10,0)$  in phase-space.
For both initializations we  initialized the spin in the state $|\hspace{-3.2pt}\downarrow\rangle$, anti-aligned with the initial effective magnetic field $\vec b(x_0,p_0,0)$. 
From the  resulting effective cavity Hamiltonian  of the system shown in  Fig.~\ref{fig:contours}a,
%(for anti-aligned spin), 
we expect  these two initializations to place the system inside and outside the frequency locking regime, respectively.

In Fig.~\ref{fig:dynamics}a, we show the  dimensionless Fourier transform of $\langle \hat x(t)\rangle$ (absolute value), $|\tilde x(\omega)|$, for the two initializations above, while Fig.~\ref{fig:dynamics}b shows a close-up of the spectrum in the vicinity of $\omega_c= \Omega /3$~\cite{fourier_transform}.
In the frequency-locked regime, $|\tilde x(\omega)|$ features  an extremely sharp peak of magnitude $\sim 10$ at $\omega = \Omega/3$. 
The two side-peaks visible in Fig.~\ref{fig:dynamics}b arise from the slow  orbit %motion %
% modulation of the cavity state discussed above:
% 
of the cavity wave-packet  around the local minimum of $\mathcal H_{\rm eff}(x,p)$ (see Sec.~\ref{sec:effective_hamiltonian}); their offset from the main peak defines the oscillation frequency of this motion. 
As is evident in Figs.~\ref{fig:dynamics}cd,  in the  regime the system has a clear, measurable subharmonic response to the driving.   
In contrast, outside the frequency-locked regime,  $|\tilde x(\omega)|$  shows a broad feature around the same value, but no well-defined peak.

When weakly coupled to an external environment (such as an electromagnetic waveguide), it may be possible to extract an output signal whose frequency spectrum shares the spectrum of $\langle x(t)\rangle$, and hence exhibits well-defined coherent oscillations at  frequency $\Omega/q$ which is evident in Fig.~\ref{fig:dynamics} . 
In this way, the  qubit-cavity system can potentially be exploited for frequency conversion.

\section{Discussion}
\label{sec:discussion}

The discovery of Floquet  time crystals sparked  a broader investigation  of discrete time-translation symmetry breaking. % phenomena. 
This work shows how such symmetry breaking can emerge as quantum frequency locking in a periodically driven spin-cavity system.
When frequency-locked, the system exhibits well-defined  oscillations  with extended  period $\T = qT$,  where $T$ denotes the driving period, and $q$ is an integer.
Quantum frequency locking moreover has remarkable consequences for the  quasienergy spectrum of a system: a large number of multiplets of  Floquet eigenstates emerge whose  quasi-energy differences are exponentially close to $n \Omega/q$ for  $n=1,\ldots,q$. 
% This multiplicity is remarkable; 
Using a semiclassical phase-space approach, we  identify two mechanisms for  frequency locking, which allow it to occur in a wide region of parameter space.
Quantum frequency locking  hence  does not require  fine-tuning, and can be reached through appropriately  {controlled but not fine-tuned} initialization of the cavity mode, for a finite range of detuning $\delta \omega =\omega_c -r\Omega/q$, and for both weak and strong qubit-cavity coupling $\eta$.
%  of parameters, and .

The frequency locking exhibited by the qubit-cavity system  is of fundamentally different nature than, e.g., time-crystalline behavior in  spin chains (see, e.g., Refs.~\onlinecite{Khemani_2016,Else_2016b}).
In the latter setting, time-translation symmetry breaking is also manifested in a large degeneracy of period-doubled Floquet eigenstates. 
However, for these systems, period multiplication  emerges from the many-body nature of the system, and  each quasienergy level in the system forms a part of a quasienergy-locked multiplet. 
In contrast,  for the qubit-cavity system,  only a finite (nonzero) number of quasienergy levels form multiplets. 
% However, the 

We expect that the nontrivial fixed points of the stroboscopic motion generated by the semiclassical effective Hamiltonian $\mathcal H_{\rm eff}(x,p)$  remain stable in the presence of weak dissipation in the cavity, as would be the case if the radiation is allowed to leak out. 
In this case  the frequency locking effect  could be used for extracting  an output signal whose frequency  is given by  a rational fraction of the drive, thus achieving frequency conversion. 
This offers an interesting direction for future studies. 
%0

The driven spin-cavity system we considered
is perhaps one of the simplest  systems that exhibits quantum frequency locking. %period multiplication. 
This generic  class of models %arise in 
can describe a diverse range of settings and physical systems, such as, e.g.,  Rydberg atoms in optical cavities and qubits in contact with microwave modes. 
 Due to the simplicity of the model, and the many suitable experimental platforms, we  expect that the qubit-cavity model  forms a convenient and versatile platform for studying the breakdown of discrete time-translation symmetry.
At  strong coupling $\eta$, frequency locking  moreover coexists with the topological energy-pumping regime that was analyzed in Ref.~\onlinecite{Nathan_2019}.  
{Thus, the relatively simple and experimentally accessible model of a driven qubit-cavity system supports several  distinct, highly nontrivial non-equilibrium phenomena. 
The simplicity of the platform, and the interplay of these nontrivial phenomena makes the driven qubit-cavity system   an interesting subject for future experimental and theoretical studies.}

{\it Acknowledgements ---} 
IM was supported by the Materials Sciences and Engineering Division, Basic Energy Sciences, Office of Science, U.S. Dept. of Energy.
FN and MR are grateful to the Villum Foundation and the European Research Council (ERC)
under the European Union Horizon 2020 Research and
Innovation Programme (Grant Agreement No. 678862) for support. GR is grateful for NSF DMR grant number 1839271.
GR is also grateful to the U.S. Department of Energy, Office of Science, Basic Energy Sciences under Award  DE-SC0019166. NSF and DOE supported GR's time commitment to the project in equal shares.

\bibliography{snake_footnotes.bib,Snake_Bibliography.bib}

\begin{thebibliography}{87}
\bibitem [{fre()}]{frequency_tuning}%
  \BibitemOpen
  \href@noop {} {}\bibinfo {note} {In Fig.~\ref{fig:Fig1}c, quantum frequency
  locking occurs for a nonzero range of driving frequencies for any given value
  of $\eta$. Hence we expect that it is enough to have good control over a
  single parameter (such as the driving frequency) to reach a regime where
  quantum frequency locked occurs. We expect this to be achievable in
  experiments.}\BibitemShut {Stop}%
\bibitem [{x_p()}]{x_p_normalization}%
  \BibitemOpen
  \href@noop {} {}\bibinfo {note} {Specifically, $x$ and $p$ are normalized
  such that $\frac{1}{2}(\hat x^2 +\hat p^2) $ gives the number of photons in
  the cavity mode.}\BibitemShut {Stop}%
\bibitem [{flo()}]{floquet_propagation}%
  \BibitemOpen
  \href@noop {} {}\bibinfo {note} {To see this, note that $\hat{U}(kT) =
  \hat{U}(T)^k$}\BibitemShut {NoStop}%
\bibitem [{oth()}]{other_families}%
  \BibitemOpen
  \href@noop {} {}\bibinfo {note} {Note that for any quasienergy-locked system
  there exists an infinite number of ways to construct a family of orthogonal
  states $|\chi_1\left.\!\!\right>,\ldots |\chi_q\left.\!\!\right>$ that
  satisfy Eqs.~(\ref{eq:qel_fes_form})~and~(\ref{eq:chi_relation}). Such
  distinct families can be obtained by modifying the (arbitrary) phase factor
  in front of each eigenstate $|\psi_\ell\rangle$ in the expression for
  $|\chi_k\rangle$ below Eq.~(\ref{eq:qel_fes_form}). However, for the
  qubit-cavity system we consider in this work, there exists a unique way to
  assign phase factors to the Floquet eigenstates
  $\{|\psi_\ell\left.\!\!\right>\}$ (up to an overall common phase factor),
  such that each $|\chi_k\left.\!\!\right>$ has support only in a particular
  well of $\mathcal H_{\rm eff}(x,p)$ in phase space. We expect analogous
  results hold for other realizations of quasienergy locking.}\BibitemShut
  {Stop}%
\bibitem [{qe_()}]{qe_correspondence}%
  \BibitemOpen
  \href@noop {} {}\bibinfo {note} {This follows since the Floquet operator in
  the rotating frame, ${ {\tilde U}}(qT)$, is given by $[\hat U(T)]^q$, where
  $\hat U(T)$ denotes the Floquet operator in the lab frame (recall that $
  {\hat U}(t)=\hat U_0(t) \tilde U(t)$, while $\hat U_0(qT) = 1$).}\BibitemShut
  {Stop}%
\bibitem [{mor()}]{more_families}%
  \BibitemOpen
  \href@noop {} {}\bibinfo {note} {Note that the same ring of potential wells
  can support multiple families of quasienergy-locked eigenstates if an
  individual potential well in the ring supports multiple bound states, as
  demonstrated in Appendix~\ref{app:other_wigners}.}\BibitemShut {Stop}%
\bibitem [{app()}]{approximate_rotation}%
  \BibitemOpen
  \href@noop {} {}\bibinfo {note} {Since $\hat U(T) = \hat U_0(T) \tilde U(T)$,
  and $\tilde U(T)$ does not transfer $|\chi_k\rangle$ out of well $k$ in the
  frequency locked regime, $\hat U(T)$ approximately generates a phase space
  rotation by $2\pi$. Moreover note that $\hat U (T)$ by construction is a
  symmetry of the exact effective Hamiltonian in the rotating frame, $\tilde
  H_{\rm eff} \equiv \frac{i}{qT} \log[\tilde U(qT)]$, regardless of any
  approximations.}\BibitemShut {Stop}%
\bibitem [{fou()}]{fourier_transform}%
  \BibitemOpen
  \href@noop {} {}\bibinfo {note} {Specifically, $\tilde x(\omega) \equiv
  \lim_{\tau \to \infty} \frac{1}{\tau} \intop _{0}^{\tau} {\rm d}t\,
  e^{i\omega t} x(t)$. The data in Fig.~\ref{fig:dynamics}cd are computed using
  $\tau = 3000 T$.}\BibitemShut {Stop}%
\bibitem [{dio()}]{diophantine}%
  \BibitemOpen
  \href@noop {} {}\bibinfo {note} {Specifically, the inherent structure of
  $H_F$ implies that terms in Eq.~(\ref{eqa:h_eff}) can only be nonzero if
  $r(m-n) = lq$ for some integer $l$. Since $q/r$ is an irreducible fraction,
  this can only be true if $l=rk$ for some integer $k$. As a result, sites $m$
  and $n$ are coupled only when $m-n=kq$.}\BibitemShut {Stop}%
\bibitem [{eig()}]{eigenstate_result}%
  \BibitemOpen
  \href@noop {} {}\bibinfo {note} {To see this, note that
  $\left\lVert\right.(\hat U(T)-e^{-i\varepsilon
  T})|\psi\rangle\left.\right\lVert =x $ implies that the smallest eigenvalue
  of $U(T)-e^{-i\varepsilon T}$ must have absolute value less than $x$.
  Moreover, we note that the component of $|\psi\rangle$ orthogonal to the
  corresponding eigenstate must be of order $x/\delta \varepsilon_n T$, where
  $\delta \varepsilon_n \sim (R^2 T)^{-1}$ was defined below
  Eq.~(\ref{eq:fourier_ft}).}\BibitemShut {Stop}%
\end{thebibliography}

\appendix
\section{Photon lattice picture of frequency locking }
\label{app:photon_lattice}
\begin{figure}
\includegraphics[width=0.99\columnwidth]{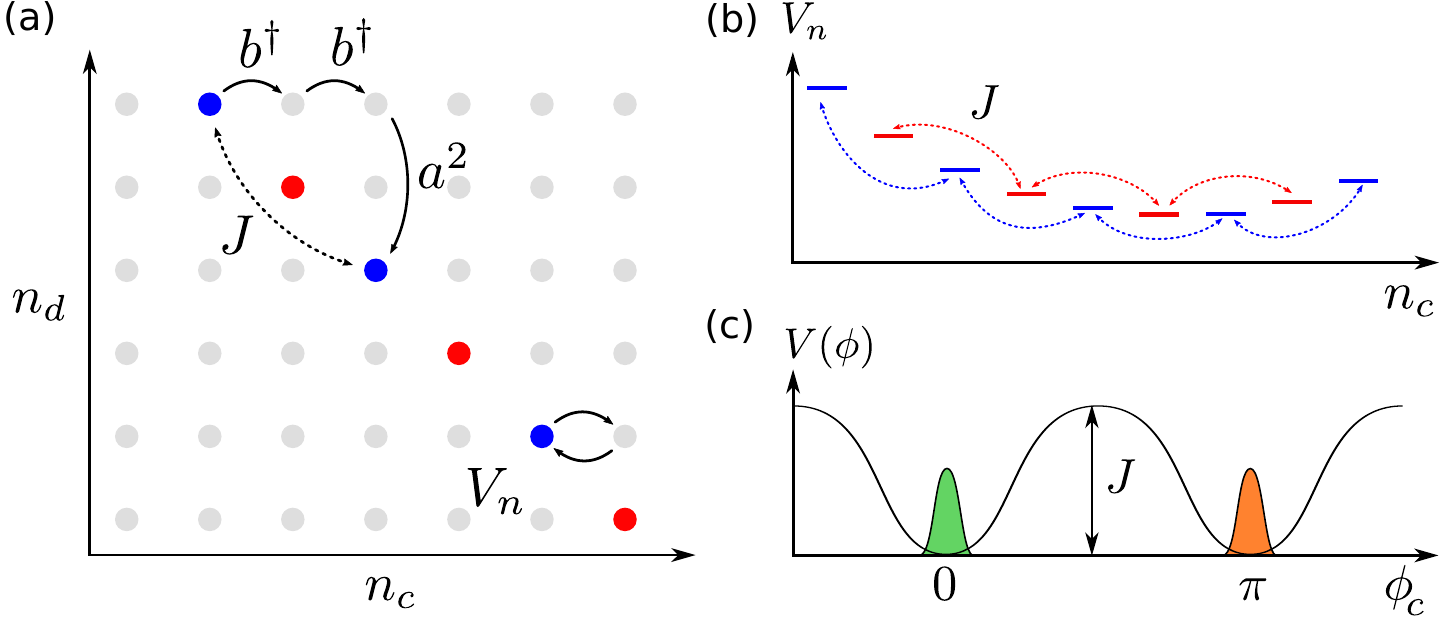}
\caption{
%{\bf is the figure correct? Shouldn't the manifold be 1 photon drive translates to 2 photons in the cavity? so a 2:1 ratio no the blue red line? }
Photon lattice representation of frequency locking when drive $\Omega$ is close to  $2 \omega_c$.
(a) Photon lattice of the system  (see main text for details). 
Red and blue  indicates the    chain of sites along which the eigenstates of $H_F$   have their primary support when $\Omega \approx 2\omega_c$. 
Arrows indicate examples of resonant virtual processes contributing to the second nearest-neighbour tunneling  $J$ and on-site potential $V_n$ of the effective tight-binding Hamiltonians of  the chain. 
Different colors emphasize the decoupling into  independent sublattices.
(b) Schematic depiction of a tight-binding Hamiltonian for the chains, with sites and on-site potential shown % and second-nearest neighbour hopping indicated
 as a function of $n_c$. 
(c) Energy profile as a function of the oscillator phase  $ \phi_c$ (variable conjugate to $n_c$), for the oscillator states close to the minimum of $V_n$.
%Here $V_\phi$ denotes the potential energy of the system as a function of $\phi_c$. 
Green and orange are approximate bound states of the effective Hamiltonian near $\phi_c=0$ and $\pi$, % when the  tunneling $J$ is sufficiently strong. 
each a superposition of ``red" and ``blue" chain states in (b).
Up to exponentially weak tunneling correction, these  bound states are also Floquet eigenstates, corresponding to the semiclassical states with the oscillator phases  locked to $0$ or $\pi$ respectively. % when the barrier, where the phase is locked to either $0$ or $\pi$. % (the phase difference between the drive and cavity) is approximately identical to the Hamiltonian of a free particle in a $\pi$-periodic cosine potential. 
%%For sufficienly large barrier width, the effective Hamiltonian support bound eigenstates confined near $ \phi = 0$ or $\pi$ (orange and green). These eigenstates hence correspond to  motion.
} 
\label{fig:lattice}
\end{figure}
In this appendix, we present a complementary perspective of frequency locking, based on the photon lattice picture of periodically driven  systems. % in the picture of the photon latice. 
The approach is used to analyze %the emergence of
frequency locking of the  the  qubit-cavity model 
in the limit of small anharmonicity $\eta$ and detuning. 
%For concreteness, 
To demonstrate the emergence of frequency locking, we  consider  the case where the driving frequency is close to a rational multiple of the cavity frequency, $\Omega \approx q \omega_c/r$, where $q$ and $r$ are integers. % %, in the limit where $\eta $ is small. 
We analyze the model as a periodically driven system with driving period $\T =  2\pi q/\Omega$ [recall that  $H(t)=H(t+T)$ implies $H(t)=H(t+\T)$]. 

%As a first step, we analyze the static photon lattice model corresponding to the $T_p$-period driven system. 
For a periodically driven system with driving period $\T$, the photon lattice Hilbert space is spanned by the orthonormal basis $|i ,n_d\rrangle = |i \rangle \otimes |n_d\rangle$, where $i $ indexes the basis states of the original problem, while  $n_d\in \mathbb Z$ can be seen as a lattice index, and  heuristically counts  the number of drive photons with energy $2\pi/\T$~\cite{Shirley_1965}. 
%For the case we study, where the period is artificially extended to $pT$, the number $n_d$ counts the number of {\it fractional} drive photons. 
The extended Hilbert space Hamiltonian reads 
$ 
H _F= \frac{2\pi}{\T} \hat  n_d  + \sum_{z,w} H^z _{ij } |i ,w+z\rrangle \llangle j ,w|,\label{eq:HFp}
$
where $\hat n_d|i,n\rrangle = n|i ,n\rrangle$, and $H_{ij  }^z$ denotes the Fourier coefficients of $H_{ij}(t)$ (as a  $\T$-periodic function of time). 
One can  verify that the eigenstates  of $H_F$, $|\psi_n\rrangle = \sum_{i,z}\psi^n_{i ,z}|i ,z\rrangle$, are related to the Floquet eigenstates of $H(t)$ as follows:
\be
|\psi_n\rangle = \sum_{i ,z}  \psi_{i,z}^n|i \rangle.
\label{eq:floquet_correspondence}
\ee
The  quasienergy of the state $|\psi_n\rangle$ is related to the corresponding energy as $\varepsilon _n = E_n \, (\mod 2\pi/qT)$. 
Note that each Floquet eigenstate of $H(t)$ corresponds to  an infinite family of eigenstates of $H_F$ due to the symmetry 
$ 
[\hat a,H_F]=\frac{2\pi}{\T},
%\label{eq:h_f_symmetry}
$
 where % $\hat a$ is the translation operator in photon space: 
 $\hat a|i ,n_d \rrangle = |i ,n_d-1\rrangle$. 
As a result, if $|\psi_n\rrangle$ is an eigenstate of $H_F$ with energy $E_n$, $\hat a|\psi_n\rrangle$ is also an eigenstate of $H_F$, with energy $E_n - 2\pi/(qT)$. Both eigenstates correspond to the same Floquet eigenstate through % the relation below Eq.~\eqref{eq:HFp}. 
Eq.~\eqref{eq:floquet_correspondence}.

%In the case we study, where the driving period  $\tau$ is artificially extended to $T_p = pT$, the index $n_d$  counts the {\it fractional} drive photons with energy $2\pi /T_p = \Omega / p$. 
To find $H_F$ for the  qubit-cavity  system, we recall that the Hilbert space of the system is spanned by the states $|\alpha, n_c \rangle$, where $n_c=0,1,\ldots $ counts the  number of cavity photons, while $\alpha =1,2$ denotes the state of the qubit. 
Hence, we can label the basis states for the extended Hilbert space $|\alpha ,n_c,n_d\rrangle = |\alpha ,n_c\rangle\otimes |n_d\rangle$. 
We write 
$ 
H_F = V +K
\label{eq:h_f_def}
$
where $V$ and $K$ denote the diagonal and off-diagonal components in the basis above.
To find $V$ and $K$, we recall from Eqs.~\eqref{eq:hamiltonian_1}-\eqref{eq:j_sc_def} in the main text that   the Hamiltonian $H(t)$ oscillates monochromatically with period $T$.  Therefore, $H_{ij  }^z$ % the terms in Eq. (\ref{eq:HFp}) are
is nonzero only when $z$ is an integer multiple of $q$.
Equivalently, the above-introduced photon number shift operators $\hat a$ and $\hat a^\dagger$ only appear in powers of $q$ in the expression for $H_F$.
Using the expression for $H(t)$ in Eqs.~\eqref{eq:hamiltonian_1}-\eqref{eq:j_sc_def}, we find 
\begin{eqnarray}
V &=&\hat n_c \omega _c + \frac{\Omega}{q} \hat n_d +\eta  \sigma_x B_0  \label{eq:h_f_components}\\ 
K &=&  \frac{\eta A_d}{2}  \left(\hat a^q [\sigma_z -i \sigma_x]+ \hat a^{\dagger q}[\sigma_z+i\sigma_x]\right)+\eta ( \hat b \sigma^+ + \hat b^\dagger \sigma^{-}). \notag 
\end{eqnarray}

In the same way as for example in Refs.~\onlinecite{Martin_2017,Nathan_2019}, we can see  $H_F$  as describing a $2D$ square lattice tight-binding model where $(n_c,n_d)$  denotes the site index in the ``photon'' lattice, and $\alpha $ denotes the orbital index. 
%We refer to this  $(n_c,n_d)$  as the {\it photon lattice}.
The sites in the photon lattice are coupled by the term $K$, and are subject to the on-site potential energy term $V$. 
Note that, by construction, the Hamiltonian $H_F =V+K$  only couples sites $(n_c,n_d)$ in the photon lattice separated by a distance $q$ in the second coordinate.
%This structure is a consequence of the additional symmetry $H(t)=H(t+T)$ which artificially arises from choosing the ``long'' driving period $T_p$. 

In the limit $\eta \to 0$, the term $V$ will generally dominate, and the eigenstates of $H_F$ are localized on individual sites in the photon lattice. The eigenstates  are given by $|\Psi_{mn}^\pm \rrangle  \approx\frac{1}{\sqrt{2}} (|1,m,n\rrangle \pm|2,m,n\rrangle) $, with energies
\be 
E_{mn}^{\pm} = m\omega_c+\frac{ n\Omega}{q} \pm \eta B_0.
\ee
These solutions are trivial, and hence  typically there is no frequency locking in the small-$\eta$ limit.
However, when the driving frequency  $\Omega$ is sufficiently close to $\omega _cq  /r$, 
% there is an exception to the above result.
% In this case, 
the ``potential energies'' on sites $(n_c,n_d)$ and $(n_c+1,n_d-r)$  can be close enough  that 
%the term 
$K$ couples these sites resonantly through a high-order virtual processes (recall that the energy step size on the drive lattice as defined above is $\Omega/q$, while on the cavity lattice it is $\omega_c$). %(recall, that the second coordinate counts the {\em fractional} drive photons with frequency $\Omega/p$, see Fig.~\ref{fig:lattice}a).  
As a result, each  eigenstate of $H_F$  may  extend along a  chain of  sites $(k ,b -r k )$ %{\bf Should this be $(k\cdot q,b-r k)$?}
for $k=0,1,\ldots$, as depicted in Fig.~\ref{fig:lattice}a for the case $q/r=2$. 
Due to the symmetry of $H_F$ described below Eq.~\eqref{eq:floquet_correspondence}, there is just one independent chain ($b=0$),  while the remaining chains are related by shifts in $n_d$.

Each chain is  subject to an effective Hamiltonian which arises from the high-order virtual processes. 
This Hamiltonian takes the form
\be 
H_{\rm eff} =  \sum_{m,n} |\alpha ,m,-rm\rrangle\llangle \beta , n,-rn|H^{\alpha \beta }_{mn}.
\label{eqa:h_eff}
\ee
Here the  matrix elements $H_{mn}^{\alpha \beta }$ can  in principle be calculated  analytically from perturbation theory in $\eta$. 
%To demonstrate how frequency locking may arise in the model, we consider the general properties of the effective Hamiltonian $H_{\rm eff}$ above. 
Since $H_F$ by construction only couples sites $(n_c,n_d)$ in the photon lattice  separated by a distance $q$ in the second coordinate, 
the terms off-diagonal in photon number basis can  only be nonzero  when $m-n=kq$ for some integer $k$\cite{diophantine}.
This coupling arises from virtual   processes where $kp$  cavity photons are emitted, and $kq$ full  drive photons (with frequency $\Omega$) are absorbed, or vice versa.

The above considerations show that the 1D chain model above itself  separates into $q$  decoupled sublattices, distinguished by the value of $n_c \, (\mod q)$. 
The  tunneling coefficient $H_{m,m+kq}^{\alpha \beta }$  arises from a $k(q+r)$-th order virtual  process (see Fig.~\ref{fig:lattice}a), and hence scales  as $\eta ^{k(q+r)}$. 
Thus, only the $k=1$ term is relevant in the $\eta \to 0$ limit.    % and arises from $(p+q)$th order
Following this discussion, we conclude that $H_{mn}^{\alpha \beta }$ takes the form 
\be 
%H^{\alpha \beta }_{mn} = V^{\alpha \beta }_n \delta_{mn} + \frac{1}{2}(J^{\alpha \beta }_n \delta_{m,n+p} + J^{ *\beta \alpha  }_n\delta_{n+p,m})
H_{mn} = V_n \delta_{mn} + \frac{1}{2}(J_n \delta_{m,n+q} + J^\dagger_n\delta_{n+q,m})
\label{eq:h_eff_form}
\ee
where $V_{n}$ and $J_n$ are $2\times 2$ matrices acting on the Hilbert space of the  qubit, and we suppressed the qubit indices $\alpha ,\beta $ for brevity.
The term $V_n$ has contributions from the static field $B_0$, from the finite detuning $\delta \omega$, and from even-order ``closed'' virtual processes, while the origin of the term $J_n$ was discussed in the above. % that act exclusively to renormalize the on-site energies, such as the ones depicted in the bottom of  Fig.~\ref{fig:lattice}a. 

%The above line of arguments is schematically illustrated in Fig.~\ref{fig:lattice}a, for the case $p/q = 2 $. % $p/q=2$. 
%Here, we depict  the resulting    $2$ disconnected chains (consisting of sites  $(2k,-2k)$ and $(1+2k,-2k-1)$, respectively) by black and red sites in the photon lattice. %in the photon lattice, namely the , . % (corresponding to $
%%(w,z)=(1,0)$ and $(0,0)$, respectively. 
%%In the limit where $\Omega \approx 2 \omega_c$, the eigenstates of $H_F$ either have support on the red or the black chain. 
%In the limit of small $\eta$,  the two chains are disconnected from the remaining sites due to the term $V$ in $H_F$, while sites within each  chain are  coupled other sites in the chain through $3$rd order virtual processes, where $2$ cavity photons are emitted, and two driving photons are absorbed, or vice versa.

While it is straightforward to analytically compute the terms $V_n$ and $J_n$ above through  perturbation theory in $\eta$, such an analysis is beyond the scope of this paper. 
Instead, below we infer the emergence of frequency locking from a more qualitative discussion of the effective Hamiltonian above. 
We consider the case of a spinless model, where the coefficients $V_n$ and $J_n$ in Eq.~\eqref{eq:h_eff_form} are scalars. 
Such a Hamiltonian emerges when the above line of arguments is applied to a periodically-driven anharmonic oscillator, such as considered in Ref.~\onlinecite{Zhang_2017}. 
We expect that the ``spinful'' model  arising from the qubit-cavity system can be analyzed in a similar way.

To see how frequency locking arises in the spinless model, we  note that for a finite range of detuning $\delta \omega = \omega- r\Omega/q$ the net potential energy $\tilde V_ n = V_n + |J_n|$  may have a nontrivial minimum as a function of $n$, as schematically illustrated in Fig.~\ref{fig:lattice}c (the case of a maximum is similar). 
Near the minimum $n_0$ of $\tilde V_n$,  to lowest order in $n-n_0$, $H$ takes the form
\be
H_{mn} \approx  \frac{k}{2} (n-n_0)^2\delta_{mn}  + \frac{J}{2} (\delta_{m,n+q}+\delta_{m,n-q}-2),
\ee
where $J=J_{n_0}$, and the ``spring constant'' $k$ can be computed from Taylor expanding  $\tilde V_n$ around $n=n_0$.
It is illuminating to express the Hamiltonian above in terms of the variable $  \phi$ conjugate to  $n-n_0$: 
\be 
 H_{\rm eff} = - \frac{1}{2m_{\rm eff}}\partial _\phi^2   + J(\cos (q   \phi)-1),
\label{eq:phase_hamiltonian}\ee
where $m_{\rm eff} = 1/k$. 
Physically, since the index $n$ measures the value of $ \hat n_c$ (i.e. the number of cavity mode photons) up to a constant shift by $n_0$ [see Eq.~\eqref{eqa:h_eff}],  $  \phi$  measures the phase of the cavity mode.
Thus, when $ n_c  \approx n_0$ the effective Hamiltonian  for the phase of the cavity mode describes the Hamiltonian of a free particle in a cosine potential $V(\phi)$  with well spacing $2\pi/q$ and depth $J$, as depicted in Fig.~\ref{fig:lattice}c.

{
Importantly, when the potential well depth $J$ is sufficiently large compared to the of kinetic energy of  zero-point fluctuations   associated with the effective mass $m_{\rm eff}$,} the effective Hamiltonian above may support bound states where wave function of the system (as a function of $\phi$) is confined to one of the potential wells. 
In this state, with exponential accuracy, the phase of the oscillator $\phi$ is  locked to an integer multiple of $2\pi/q$.

We now demonstrate that these bound states can be used to construct Floquet eigenstates of the qubit-cavity model where the phase has locked to the driving field [recall that the eigenstates in the photon lattice correspond to Floquet eigenstates of the qubit-cavity system through Eq.~\eqref{eq:floquet_correspondence}]. Indeed, from the bound states $|\psi_z\rrangle$ localized in isolated potential wells $z$,  one can construct plane-wave'' combinations, $|\Psi_n \rrangle = \frac{1}{\sqrt{q}}\sum_z |\psi_z\rrangle e^{-\frac{2\pi i  z n}{q}}$. 
Due to gaussian confinement of the wavefunction in the bottom of the near-harmonic potential wells in Eq.~\eqref{eq:phase_hamiltonian}, the energy differences between these distinct combinations will be exponentially small in $\lambda^2/\xi^2$, where $\lambda = 2\pi/q$ denotes the well separation, and $\xi = (J m_{\rm eff})^{-1/4}$ denotes the scale of the phase fluctuations around the potential minimum. 

Through the correspondence between eigenstates of $H_F$  and the  Floquet eigenstates of the system, we conclude there must exist families of $q$ Floquet eigenstates, whose quasienergies differ by an integer multiple of $\Omega /q$, up to a correction $\delta \varepsilon $ exponentially small in $\lambda^2 /\xi^2 \sim \sqrt{J m_{\rm eff}} /q^2$.  
This is in agreement with the main text, where we indeed found  multiplets of   Floquet eigenstates  with exponentially close quasienergies modulo $\Omega /q$.

\section{Frequency locking at other frequency ratios}
\label{app:higher_mode_locking}
\begin{figure*}
\includegraphics[width=2.1\columnwidth]{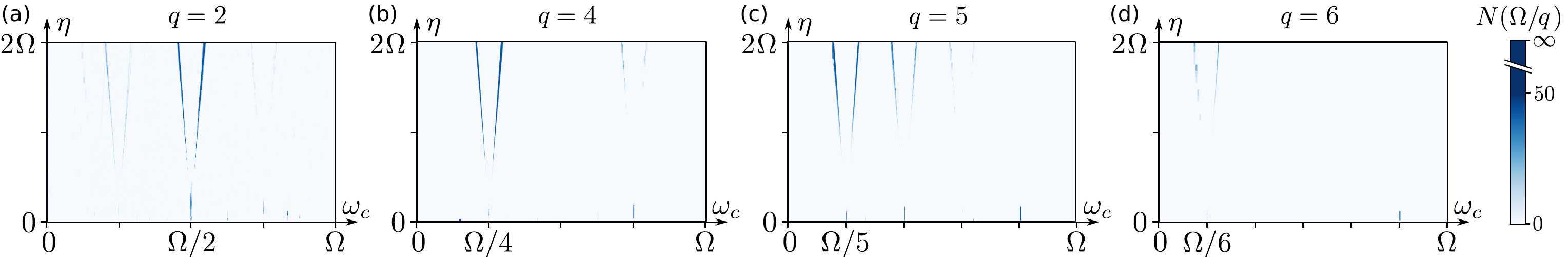}
\caption{Number of period-$q$ frequency locking Floquet eigenstates as a function of $\omega_c$ and $\eta$, for the model depicted in Fig.~\ref{fig:Fig1}c, for $q=2$  (a),  $q=4$ (b), $q=5$  (c), and  $q=6$(d).
See Appendix~\ref{app:higher_mode_locking} for further details. }% (panel (a)]}
\label{fig:higher_mode_locking}
\end{figure*}

Here we numerically demonstrate  that quantum frequency locking also occurs at ratios other than $3$. 
For the same model and realizations studied in Sec.~\ref{sec:numerics} (see Fig.~\ref{fig:Fig1}c), we counted the number of frequency locking Floquet eigenstates at period multiplication $q$ for $q=2,4,5,6$, as a function of the coupling strength $\eta$ and cavity frequency $\omega_c$. 
The frequency locking states were identified from the quasienergy level spacings (modulo $\Omega/q$),  in the same way as for the period-$3$ frequency locking states (see  Sec.~\ref{sec:numerics}). % for more details).
Our counting procedure identified a unique period multiplication for each frequency locking state,  such that period-$2$ frequency locking eigenstates were not also double-counted as a period-4 Floquet eigenstates. 
In Fig.~\ref{fig:higher_mode_locking}, we plot the obtained number of period-$q$ frequency locking states. 
%for $q=2,4,5,6$, as a function of the qubit-cavity coupling $\eta$ and cavity frequency $\omega_c$. 
(Note that a  different color scale is used compared to Fig.~\ref{fig:Fig1}, in order to heighten the contrast.)
Fig.~\ref{fig:higher_mode_locking} clearly shows the same branch structure as Fig.~\ref{fig:Fig1}c, with period-$q$ frequency locking occurring whenever $\omega_c/\Omega$ is close to $r/q$ for  integer $r$.
This demonstrates that period-$q$ frequency locking can occur for any $q$, when $\omega_c/\Omega$ is close enough to $r/q$ for some integer $r$. 

\section{Derivation of $\mathcal H_{\rm eff}$ in Floquet regime}
\label{app:heff}
\label{app:floquet_locking}
Here we identify the conditions for the  Floquet  regime, and derive the corresponding semiclassical effective  Hamiltonian $\mathcal H_{\rm eff}(x,p)$ [Eqs.~\eqref{eq:heff}~and~\eqref{eq:heff_expression_floquet}].
These results were quoted in Sec.~\ref{sec:floquet} of the main text.

\subsection{Conditions for the Floquet regime}
We begin by deriving the conditions for the Floquet regime that we quoted in Sec.~\ref{sec:floquet}, namely:
\be 
\eta \ll \theta(x,p)/\tilde T, \quad \delta \omega \ll \tilde \Omega/A_{\rm cav}
\label{eqa:floquet_conditions}
\ee
where $\theta(x,p)$ denotes the stroboscopic precession angle (see sec.~\ref{sec:floquet}), and $A_{\rm cav}\equiv \sqrt{x^2+ p^2}$ denotes the displacement amplitude of the cavity mode.

To identify the conditions for Floquet locking, we  consider the spin's dynamics [Eq.~\eqref{eq:s_eom} in the main text] for fixed $x$ and $p$. 
In this case $\vec S(t)$ evolves according to the  Schr\"odinger-type equation  % with a $3\times 3$ Hamiltonian matrix $H_q(t)$: 
\be
\partial _t S_k(t) =  -i \sum_{l}H_{kl}^{(3)}(x,p,t) \vec S_l(t),
\label{eqa:seom}\ee 
where  $H_{kl}^{(3)}(x,p,t)= -i \eta \sum_j  h_j (x,p,t) \epsilon_{jkl}$ is a $3\times 3$ Hermitian  matrix, with $ \epsilon_{jkl}$ denoting the Levi-Civita tensor. %the generators of $3d$ rotation.
 Due to the Floquet theorem, % time-periodicity of $H^{(3)}(x,p,t)$, % Floquet theorem, 
 Eq.~\eqref{eqa:seom}
 %the equation of motion above %for $\vec S(t)$
  has $3$ complex-valued orthonormal  solutions of the form $\vec S(t) = \vec n_j(t) e^{-i\varepsilon_jt}$, where   $\vec n_j(t)=\vec n_j(t+\T)$.
The antisymmetry of $H^{(3)}(t)$  %(class D in the AZ classification~\cite{AltlandZirnbauer})
implies that one of the stationary solutions is  real-valued, with  quasienergy zero. 
We identify this solution as the vector $\vec n_0(x,p,t)$ from Sec.~\ref{sec:floquet}.
Up to a prefactor, $\vec n_0(x,p,t)$  is the unique   $\T$-periodic solution to Eq.~\eqref{eq:s_eom}, with $x$ and $p$ fixed. % $\partial _t \vec S(t) = -\vec h(x,p,t)\times \vec S(t)$.
The remaining  two  orthogonal solutions  are  related to each other by Hermitian conjugation, and have quasienergies $\pm \theta(x,p)/\tilde T$. 
The above properties imply that the effective Hamiltonian associated with $H^{(3)}(x,p,t)$ is given by 
\be
H_{\rm eff}^{\rm spin}(x,p) =[ {\theta(x,p)}/{\tilde T}] \vec a(x,p)\cdot \vec S,
\label{eq:effective_bloch_equation}
\ee
where we used $\vec n_0(x,p,t)= \vec a(x,p)$. 
% Comparing with the expression for $H_{\rm eff}^{\rm spin}(x,p)$ in Sec.~\ref{sec:floquet_locking}, we conclude $\varepsilon _0(x,p)=\varepsilon (x,p)$. 
Thus, we have related $\vec a(x,p)$, $\vec n_0(x,p,t)$ and $\theta (x,p)$ to the Floquet states and quasienergy spectrum of the $3\times 3$ antisymmetric Hamiltonian $H^{(3)}(x,p,t)$.

As quoted in Sec.~\ref{sec:floquet} (see also Ref.~\onlinecite{Weinberg_2017}), the trajectory of the  spin $\vec S(t)$ is locked to $\vec n_0(x(t),p(t),t)$  % Floquet locking occurs 
 when the change of the stroboscopic precession axis $\vec a(x,p)=\vec n_0(x,p,0)$  due to the motion of $x$ and $p$ is adiabatic with respect to the quasienergy gap of the effective spin Hamiltonian, $\delta \varepsilon  (x,p) \equiv \frac{1}{\tilde T}\min(\theta(x,p),2\pi-2\theta(x,p))$: 
 \be 
\left| \frac{d}{dt} \vec a(x(t),p(t))\right|
  \ll \delta \varepsilon (x(t),p(t)),
  \label{eq:app_floquet_locking}
 \ee
where (with $x$, $p$, and $t$ suppressed)
\be 
\frac{d}{dt} \vec a = \dot x \partial _x\vec a   + \dot p \partial _p\vec a. 
\label{eq:delta_n_exp}
\ee
To identify the conditions under which Eq.~\eqref{eq:app_floquet_locking} holds,
we thus need to bound $|\partial _x\vec a|$ and $|\partial _p\vec a|$. 
%the $x$-  and $p$-derivatives of the stroboscopic rotation axis $\vec a(x,p)$.

%To relate $\vec a(x,p)$ to the field $\vec h(x,p,t)$,

To  bound   $\partial _x \vec a$, we consider the Floquet operator $R(x,p,\T)$, %generated by $H^{(3)}(x,p,t)$,
 where $R(x,p,t) \equiv \mathcal T e^{-i\int_0^t \!{\rm d}t'\, H^{(3)}(x,p,t')}$ denotes the time-evolution operator generated by $H^{(3)}(x,p,t)$.
 Due to the antisymmetry of $H^{(3)}(x,p,t)$, $R(x,p,t)$ is an orthogonal matrix.
Below we show through standard perturbative arguments that % (see below) show that  
\be 
|\partial _x \vec a |\leq  \frac{2\|\partial _x R(\T)\|}{ \pi \delta \varepsilon (x,p)\T },
\label{eq:perturbative_result}
\ee
where $\norm{\cdot}$ refers to the spectral norm, and $\delta \varepsilon (x,p)$ was defined above Eq.~\eqref{eq:app_floquet_locking}.

To prove Eq.~\eqref{eq:perturbative_result}, 
 note that, since $ R(\T) \vec a = \vec a$. 
Using the shorthand $R\equiv R(\tilde T)$ here and below, we hence have $\partial _x (R\vec a) = \partial _x \vec a$. %, where we 
Applying the chain rule, we also find % have 
$
\partial _x (R \vec a) = (\partial _x R) \vec a + R\partial _x\vec a$. 
Equating  these two expressions, we obtain
\be
(\partial _xR) \vec a =  (1-R)\partial _x \vec a.
\ee
From the spectral decomposition of $R$, we have, for any vector $\vec v$, $|(1-R)\vec v|\geq  |e^{-i\delta\varepsilon  T}-1||\vec v|$. 
Using this result along with $|e^{-i\alpha }-1|\geq \alpha \pi/2$  (for $\alpha \leq pi$) in the equation above, we obtain $   |\partial _xR \vec a|  \leq  \delta \varepsilon \T \pi |\partial _x \vec a|/2$.
Using $|\partial _x R \vec a| \leq \norm{\partial _x R}$, Eq.~\eqref{eq:perturbative_result} follows.

Using the chain rule and the triangle inequality, one can verify $\|\partial _x R(x,p,\T)\| \leq \int_0^{\T}{\rm dt}\|\partial _x H ^{(3)}(x,p,t)\|$. 
Since $\|\partial _x H _q(x,p,t)\| \leq \eta |\partial _x \vec h| = \eta$, we then find 
$
\|{\partial _x R(x,p,\T)}\| \leq \eta \T.
$
Thus, we conclude 
\be 
|\partial _x \vec a| \leq \frac{2\eta}{\pi \delta \varepsilon(x,p) }.
\label{eq:dx_n_bound}
\ee
The same bound holds for $|\partial _p \vec a|$ by similar arguments. 
Using the above result along with $\dot x , \dot p \leq  \eta + \delta \omega A_{\rm cav}$ [see Eqs.~\eqref{eq:x_eom}-\eqref{eq:p_eom}] and the triangle inequality, we finally obtain 
\be 
\left|\frac{d}{dt} \vec a\right| \leq   \frac{4\eta}{\pi \delta \varepsilon(x,p) }(\eta + A_{\rm cav} \delta \omega).
\ee
Hence, using $4/\pi \sim 1$, the condition $|d\vec a/dt|\ll \delta \varepsilon $ is satisfied when 
 \be
\delta  \varepsilon ^2 \gg \eta^2 , \, \eta \delta \omega A_{\rm cav}.
\ee
Using $\delta \varepsilon \sim \theta(x,p)/\tilde T$, we see that the first condition above is equivalent to $\eta \ll \theta /\tilde T $, which is the first condition in Eq.~\eqref{eqa:floquet_conditions}. % quoted in Sec.~\ref{sec:floquet}. 
When $\delta \varepsilon \sim \theta(x,p)/\tilde T$, the second condition above ($\delta \varepsilon  ^2 \gg \eta \delta \omega A_{\rm cav}$) is met if $\delta \omega \ll \theta /\tilde T A_{\rm cav} $; this is the second condition in Eq.~\eqref{eqa:floquet_conditions}. % Sec.~\ref{sec:floquet} of the main text.
Thus the Floquet regime arises when the two conditions in Eq.~\eqref{eqa:floquet_conditions} are satisfied. This was what we wanted to show. %quoted in Sec.~\ref{sec:floquet} are satisfied. 

% The latter two conditions in Eq.~\eqref{eq:floquet_locking_conditions} come from requiring that $\tilde x$ and $\tilde p$ remain stationary within a driving period. 
% This condition allows integrating out the time-dependence, similar to Sec.~\ref{sec:adiabatic_locking} in the main text.
% The quasistationarity condition is 
%  $\Delta x(t),\, \Delta p(t) \ll A_{\rm cav}$  for $t<\T $. % (see Sec.~\ref{sec:adiabatic_locking}). 
% %
% %obtain a time-independent  effective Hamiltonian for the cavity, 
% %From Eqs.~\eqref{eq:floquet_locking_1}
% From the same line of arguments as in % $|\vec v_s| \leq \delta \omega A + \eta $, along with the arguments from 
% Sec.~\ref{sec:adiabatic_locking} (using $|\dot x |, |\dot p| \leq \delta \omega A + \eta $), we  find that this condition is satisfied when $\delta \omega \T\ll 1 $ and $\eta \T \ll  A$. 
% These are the two last conditions quoted in Eq.~\eqref{eq:floquet_locking_conditions}.

\subsection{Derivation of effective Hamiltonian}
We now derive the effective Hamiltonian in Eqs.~\eqref{eq:heff}~and~\eqref{eq:heff_expression_floquet}.
In the Floquet  regime, whose conditions were identified above, the discussion in Sec.~\ref{sec:floquet} implies that  when the spin is initially aligned or anti-aligned with the stroboscopic precession axis, $\vec S(0) = \pm \vec a(x(0),p(0))$, the   resulting evolution  satisfies $\vec S(t) = \pm \vec n_0(x,p,t)$.
% , when the spin is initially aligned or anti-aligned with the stroboscopic precession axis: $\vec S(0) = \pm \vec a(x(0),p(0))$.
Using this in Eqs.~\eqref{eq:x_eom}~and~\eqref{eq:p_eom}, we obtain
\be
\dot x = v_x (x,p,t), \quad \dot p = -v_p(x,p,t),
\label{eq:floquet_locking_1}
\ee
where, for $s=x,p$, 
\begin{eqnarray}
v_x (x,p,t) &\equiv& \delta \omega p  \pm \eta  \partial _p\vec h(t)\cdot \vec n_0(x,p,t) , \\
v_p (x,p,t) &\equiv& \delta \omega x  \pm \eta  \partial _x\vec h(t)\cdot \vec n_0(x,p,t) . 
\label{eq:floquet_locking_2}
%\label{eq:v_def}
\end{eqnarray}

When the conditions for the Floquet regime [Eq.~\eqref{eqa:floquet_conditions}] are satisfied, the stroboscopic motion  of the cavity mode is adiabatic; as a result, it can effectively be  assumed stationary within a driving period. 
Hence, using the same arguments as in Sec.~\ref{sec:adiabatic} (see also Ref.~\cite{Oteo_1991}), we may integrate out the time-dependence of $v_x(x,p,t)$, obtaining
\be
\dot x \approx \bar v_x (x,p),\quad \dot p \approx  \bar v_p (x,p),
\ee
where, for $s=x,p$, 
\be
\bar v_s(x,p) = \delta \omega s \pm \frac{\eta}{\T } \int_0^{\T}\!\!\! {\rm d}t \,  \partial _s\vec h(x,p,t)\cdot \vec n_0(x,p,t) .
\label{eq:v_s_def_app}
\ee

In the remainder of this subsection, we seek to show that $\bar v_x = \partial _p \mathcal H_{\rm eff}(x,p)$ and $\bar v_p = \partial _x \mathcal H_{\rm eff}(x,p)$, where $\mathcal H_{\rm eff}(x,p)$ is given in Eqs.~\eqref{eq:heff}~and~\eqref{eq:heff_expression_floquet} of the main text.
To do this, we first use the chain rule to rewrite the integrand in the second term above for $s=p$ (the case $s=x$ follows analogously) 
\begin{align}
\partial _p \vec h \cdot \vec n_0 = \partial _p \left[\vec h\cdot \vec n_0\right] - \vec h \cdot \partial _p\vec n_0,
\label{eq:v_p_exp_1}\end{align}
where we suppressed the above quantities' dependence on $x$, $p$, and $t$ for brevity. 
We now consider the last term above. %, which we refer to as $A$. 
Since $\vec n_0$ obeys the Bloch equation [Eq.~\eqref{eq:s_eom}] $\partial _t \vec n_0 = -\eta \vec h \times \vec n_0$, we may write 
\be 
\vec h = \vec n_0 (\vec h \cdot \vec n_0) - \frac{1}{\eta} \vec n_0 \times \partial _t\vec n _0.
\ee
This result can be proven by directly inserting $\partial _t \vec n_0 = -\eta \vec h \times \vec n_0$ into the above, and using the cross product identity $\vec a\times(\vec b\times \vec c ) = \vec b(\vec a\cdot\vec  c ) -\vec c(\vec a\cdot \vec b)$ along with $\vec n_0 \cdot \vec n_0 = 1$. 
Using the above result along with $\vec n_0 \cdot \partial _p \vec n_0 =0$  (recall that $\vec n_0$ is normalized), we obtain 
\be
 \vec h \cdot \partial _p\vec n_0 = -\frac{1}{\eta}(\vec n_0 \times \partial _t \vec n_0)\cdot \partial _p \vec n_0.
\ee
Using  $(\vec a \times \vec b)\cdot \vec c = \vec b \cdot (\vec c \times \vec a)$,  and substituting into Eq.~\eqref{eq:v_p_exp_1}, we find  
\begin{align}
\partial _p \vec h \cdot \vec n_0 =  \partial _p \left[\vec h\cdot \vec n_0\right] +\vec n_0 \cdot ( \partial _t \vec n_0\times  \partial _p\vec n_0).
\label{eq:v_p_exp_1}
\notag
\end{align}
%
%\be
% \vec h \cdot \partial _p\vec n_0 = -\frac{1}{\eta}\vec n_0 \cdot ( \partial _t \vec n_0\times \partial _p \vec n_0) .
%\ee
We identify the second term as the 
%$\frac{1}{\T} \int_0^{\T}\!{\rm d}t\, F_{pt}(x,p,t)$, where %$T_l(x,p) = \frac{1}{\T} \int_0^{\T} F_{pt}(x,p)$ 
%where $F_{pt}(x,p,t)$ denotes
 the $x$-Berry flux $F_x(x,p,t)$ associated with the mapping
 of $\mathbb R^3$ to the unit sphere defined by $\vec n_0(x,p,t)$.  % with respect to $p$ and $t$. 
 One can verify that $\int _0^{\T}\!{\rm d}t\, F_{x}(x,p,t) = \partial _p \gamma (x,p)$, where $\gamma (x,p)$ denotes the Berry phase associated with the loop traversed by  $\vec n_0(x,p,t)$  on the unit sphere for $0\leq t < T$. 
 % (i.e. the area on the unit sphere swept out by the closed trajectory of $\vec n_0(x,p,t)$).
Thus, we find
\begin{equation*} 
\frac{\eta}{\T } \int_0^{\T}\!\!\! {\rm d}t \,\partial _p \vec h \cdot \vec n_0  = \frac{\partial }{\partial p}\left(\frac{\eta}{\T } \int_0^{\T}\!\!\! {\rm d}t \, \vec h\cdot \vec n_0 +\frac{1}{\T}\gamma \right).
\end{equation*}
Using this in Eq.~\eqref{eq:v_s_def_app}, we obtain 
\begin{align}
\bar v_x(x,p)= \frac{\partial }{\partial p}&\bigg(\frac{\delta \omega}{2}( x^2+p^2) +\frac{1}{\T}\gamma(x,p) \notag
\\
&\pm\frac{\eta}{\T } \int_0^{\T}\!\!\! {\rm d}t \, \vec h(x,p,t)\cdot \vec n_0  (x,p,t)\bigg),
\notag
\end{align}
where restored the  dependence  on $x$ and $p$ of the quantities above. 

The final step is to show that the second term inside the parentheses above equals  $\theta(x,p)/2 \tilde T$.
To show this, we recall that the Bloch equation $\partial _t \vec S(t) = -\eta \vec h(t)\times \vec S(t)$  describes   the evolution of the  expectation value of $\hat{ \vec S}$ with the $\T$-periodic  spin-$1/2$ Hamiltonian $\hat H_s(t) = \eta \vec h(t) \cdot \vec {\hat S}$ (here we suppressed the dependence of $\vec h$ and $\hat H_s$ on $x$ and $p$). 
Noting that the stroboscopic time-evolution of $\langle \vec S\rangle$ is generated by a rotation by the angle $\varepsilon _0 T$ around the axis $\vec n_0$ [see Eq.~\eqref{eq:effective_bloch_equation}], we conclude that the Floquet operator generated by $\hat H_s(t)$ is a $2\times 2$ unitary matrix given by $\hat U_s(\T) = e^{-i\theta(x,p)\vec n_0(0) \cdot \vec{\hat S}/2}$. 
Thus, we identify $\theta(x,p)/2\tilde T $ as the (positive) quasienergy associated with $\hat H_s(t)$.

To obtain an expression for  $\varepsilon $, we note that $|\psi (t)\rangle   = e^{-i\varepsilon t}|\phi _+(t)\rangle$ solves the Schr\"odinger equation generated by $\hat H_s(t)$, where $|\phi_+ (t)\rangle$ denotes the  Floquet state of $\hat H_s(t)$  with quasienergy $\varepsilon $. 
Thus, by direct substitution, one can verify that  
\be
 \varepsilon  = \frac{i}{\T} \int _0^{\T} \!\!\!\, {\rm d}t\, \langle
  \psi (t)|\partial _t |\psi (t)\rangle - \frac{i}{T}  \int _0^{\T} \!\!\!\, {\rm d}t\, \langle  \phi_+  (t)|\partial _t |\phi_+ (t)\rangle. 
  \ee 
We have $ \langle \psi (t)|\partial _t  |\psi (t)\rangle = i \eta \vec n(t)\cdot \vec h(t)$, where $\vec n(t)$ denotes the Bloch vector of the state $|\psi(t)\rangle$, which obeys the Bloch equation  in Eq.~\eqref{eq:s_eom}.
Note that $\vec n(t)$  is also identical to the Bloch vector of $|\phi_+(t)\rangle$. 
Since $|\phi_+(t)\rangle$ is time-periodic, $\vec n(t)$ is thus a time-periodic solution  to the  Bloch equation Eq.~(\ref{eq:s_eom}), and we  identify $\vec n(t) = \vec n_0(t)$ (the sign follows from the initial conditions).   
Identifying the latter term above as the Berry phase $\gamma (x,p)$ (with a factor of $1/\T$), and restoring $x$ and $p$, we conclude 
\be
\frac{ \theta (x,p)}{2\tilde T} =\frac{\eta}{\T } \int_0^{\T}\!\!\! {\rm d}t \, \vec h(x,p,t)\cdot \vec n_0(x,p,t) +\frac{1}{\T}\gamma(x,p) .
  \ee 
  A similar result holds for $\bar v_p$. 
%   Noting that $\varepsilon (x,p)= 2\theta(x,p)$, t
  This was what we wanted to show, and concludes this Appendix. 
% the result in the main text   follows.

\section{Derivation of quasienergy locking}
\label{app:qe_locking}
%\addFN{\bf  new appendix (copied in from main text of earlier version, with a few modifications)}
%\addFN{Give heuristic version with potential wells of H effective. Put rigorous version in the Appendix.

In this Appendix, we  show how quasienergy locking arises in the frequency-locked regime of the driven qubit-cavity system, using a more rigorous line of arguments than those presented in the main text. %Sec.~\ref{app:wavepackets}).
The discussion proceeds as follows:
In Sec.~\ref{app:wavepackets},  we first  consider the  qualitative behavior of the  qubit-cavity
  system in the  regime.
In Sec.~\ref{app:floquet_states}, by reconciling this behavior with Floquet eigenstate decomposition of the time-evolution (see Sec.~\ref{sec:qel_def}), % Eq.~\eqref{eq:floquet_state_decomp}, 
we  conclude   that such nontrivial extrema of $\mathcal H_{\rm eff}(x,p)$ implies the existence of multiplets of quasienergy levels of the form in Eq.~\eqref{eq:qel} in the main text, whose corresponding Floquet states take the form in Eqs.~\eqref{eq:qel_fes_form}-\eqref{eq:chi_relation}.

\subsection{Wavepacket dynamics in $q$-fold potential wells}
\label{app:wavepackets}
To show how quasienergy  locking arises, we  consider the semiclassical effective Hamiltonian of the cavity mode in the  regime,   $\mathcal H_{\rm eff}(x,p)$ (see Sec.~\ref{sec:effective_hamiltonian}); i.e., in the Floquet or adiabatic regimes,  for parameters where 
   $\mathcal H_{\rm eff}(x,p)$ acquires  extrema at finite amplitude in phase space.
%The effective Hamiltonian  $\mathcal H_{\rm eff}(x,p)$ features extrema at finite displacement amplitude in phase space [i.e., for $(x,p)\neq (0,0)$].
As explained in Sec.~\ref{sec:qel_derivation} in the main text, $\mathcal H_{\rm eff}(x,p)$ has a built-in symmetry of discrete rotation  by   $2\pi/q$ in phase space.
As a result, each potential well of this Hamiltonian (at finite displacement amplitude) forms part of a ring of $q$ potential wells that are related to each other through phase space rotations by  $2\pi/q$.

We consider the rotating-frame time-evolution, $|\tilde \psi(t)\rangle$, of a wave packet $|\tilde \psi(0)\rangle$ initialized in one of the potential wells of $\mathcal H_{\rm eff}(x,p)$, referred to as well $0$ in the following.
For example, $|\tilde\psi(0)\rangle$ may describe a direct-product state where the cavity mode is in a coherent state whose center in phase space $(x_0,p_0)$ is located in well $0$,  while the spin is initialized along or against either $\vec h(x_0,p_0,0)$  (for the adiabatic regime) or $\vec a(x_0,p_0)$ (for the Floquet regime)
\cite{spin_initialization}.
Note that the corresponding initial state of the system in the lab frame, $| \psi(0)\rangle$, coincides with $|\tilde \psi(0)\rangle$ [see Sec.~\ref{sec:effective_hamiltonian}].
Since $|\tilde \psi(t)\rangle$ must propagate along the constant-value contours of   $\mathcal H_{\rm eff}(x,p)$, we expect that $|\tilde\psi(t)\rangle$  remains confined within well $0$   up to  the  timescale  of quantum tunneling     between the  potential wells of $\mathcal H_{\rm eff}$, $\tau$. 
  We expect the  rate of quantum tunneling,  $1/\tau$, to be exponentially suppressed in  $d/\xi$, where $d$ denotes the separation between the potential wells in phase space, and $\xi\sim 1$ denotes the scale of quantum fluctuations. 
Hence, the duration of the confinement, $\tau$, scales exponentially with $d/\xi$. 

We now consider the corresponding evolution of the system in the lab frame, $|\psi(t)\rangle$. 
We recall from Sec.~\ref{sec:effective_hamiltonian} that  $|\psi(t)\rangle$ is obtained from $|\tilde \psi(t)\rangle$ through  a phase space rotation by  $2 \pi r  t/qT$. 
Hence,  for integer $k$ where $k\ll \tau /T$,  the  support of $|\psi(kT)\rangle$ in  the phase space  (of the cavity mode)
 is confined to the potential well of $\mathcal H_{\rm eff}$ which is located at an angle $2\pi  r k /q$ from  well $0$ in phase space; we refer to this potential well  as well $k$ in the following.

\subsection{Implications for Floquet eigenstates}

\label{app:floquet_states}
Above we showed that,  in the quantum  regime,  the qubit-cavity system  supports solutions  that  remain confined to potential well $k$ of $\mathcal H_{\rm eff}$ at time $t=kT$ (for integer $k$). 
The confinement persists for times $t\ll\tau$, where $\tau$ is the exponentially long
 timescale  for tunneling between the potential wells of $\mathcal H_{\rm eff}$. 
%The above behaviour
%The existence of such solutions : %, as we explore below.
%Here 
In this section, we show how the resolvability of %such solutions % $
such a confined solution, $|\psi(t)\rangle$, in terms of Floquet eigenstates,
%(see Sec.~\ref{sec:qel_def}),  
%in terms of Floquet eigenstates we consider the implications of  such confined solution  for the Floquet eigenstates and quasienergy spectrum of the system. 
%We use that   the stroboscopic time-evolution these confined solutions can be resolved  in terms of the Floquet eigenstates and quasienergies~\cite{floquet_propagation}:
\be
|\psi(kT)\rangle = \sum_n c_n e^{-i\varepsilon _n kT}|\psi_n\rangle,
\label{eq:floquet_state_decomp}
\ee  % show how the resolvability of %such solutions % $
%such a confined solution, $|\psi(t)\rangle$, 
%in terms of Floquet eigenstates 
%[Eq.~\eqref{eq:floquet_state_decomp}],
dictates the behaviour of Floquet eigenstates and quasienergies of the system.
Specifically, %we show that
each  Floquet eigenstate which significantly overlaps with $|\psi(0)\rangle$ must form a part of a multiplet of $q$ Floquet eigenstates of the form in Eq.~\eqref{eq:qel_fes_form}~and~\eqref{eq:chi_relation}, while the  corresponding quasienergies are of the form in Eq.~\eqref{eq:qel}.

To simplify the analysis, we restrict  the cavity mode's phase space  to the region where  $|\psi(t)\rangle$ has its  support: 
we let $r_{\rm max}$ denote the maximal distance $r$   from the origin  in the cavity mode's phase space where $|\psi(t)\rangle$ has significant support, and discard all   states in the Hilbert space with more than   $ R^2$ photons of the cavity mode,  for some $R\gg r_{\rm max}$. 
This truncation %operation
effectively discards the region of phase space  located more than a distance $R$ from the origin; hence we  do not expect it  to significantly affect %the evolution of the state of the system, 
$|\psi(t)\rangle$.  
Moreover, we expect the Floquet eigenstates (and their corresponding quasienergies) to remain nearly unaffected by the truncation when they have full support  well within a distance  $R$ from the origin of  phase space (up to exponentially small corrections).
We assume $R$ can be chosen several orders of magnitude %much
smaller than $(\tau/T)^{1/5}$, while still remaining much larger than $r_{\rm max}$. 
This is safe to assume since, as we recall from Sec.~\ref{app:wavepackets},  $\tau\sim e^{d/\xi}$ where $\xi \sim 1$ is the scale of quantum fluctuations, and $d$ the distance of the potential wells from the origin. 
In contrast, $r^2_{\rm max}$  only scales quadratically with $d/\xi$.

To characterize the properties of the Floquet eigenstates of the system  with the truncated Hilbert space, we consider the stroboscopic time-evolution of the confined wavepacket, 
$|\psi(mT)\rangle$, for $m=0,1,\ldots N$ for some $N$. 
For sufficiently large $N$ (see below for specific conditions), we may use this 
to %time-evolution contains all the     information necessary for  
compute  any Floquet eigenstate $|\psi_n\rangle$ whose overlap with the initial state, $c_n\equiv \langle \psi_n|\psi(0)\rangle $, is  significant:
\be
|\psi_n\rangle = \frac{1}{ c_n N}\sum_{m=0}^N |\psi(mT)\rangle e^{im \varepsilon_n T}  +\mathcal O([ c_n N\delta \varepsilon_n  T]^{-1}).%|\delta \psi_n\rangle,
\label{eq:fourier_ft}
\ee
Here $\varepsilon _n$ denotes the quasienergy corresponding to $|\psi_n\rangle$, and  $\delta \varepsilon _n\equiv \min_{m}|\varepsilon _n-\varepsilon _m|$ denotes the distance to the nearest adjacent  quasienergy  from $\varepsilon _n$.
% %and $c_n\equiv \langle \psi_n|\psi(0)\rangle$ denotes the overlap of the initial state % $|\psi(0)\rangle$
% with $|\psi_n\rangle$.
In the above, $\mathcal O(x)$ denotes a state with norm $\lesssim |x|$. 
The above result can be verified by direct insertion of Eq.~\eqref{eq:floquet_state_decomp} into the right-hand side  above.

Since  the truncated system  has  $2R^2$ quasienergy levels, uniformly distributed over the interval between $0$ and $\Omega$,  the quasienergy spacing satisfies $\delta \varepsilon _n \sim \mathcal O ([R^2T]^{-1})$. 
We refer to the overlap coefficient $c_n$ as being significant if  $|c_n|\geq 1/\sqrt{2R^2}$ (at least one such Floquet eigenstate state should exist, due to the normalization of $|\psi(0)\rangle$ which implies $\sum_{n=1}^{2R^2} |c_n|^2 = 1 $). % of $|\psi(0)\rangle$ and  the Hilbert space dimension of $2R^2$), 
Hence, for  significantly overlapping Floquet eigenstates, the correction in Eq.~\eqref{eq:fourier_ft} is, at most, of order $R^3/N$.
We  choose $N$ to be much smaller than, but still of same magnitude as $\tau/T$. 
%but still much smaller than this number.
With this choice, $N$ is much larger than $R^3$ (due to our assumption above that $R$ could be chosen much smaller than $(\tau/T)^{1/5}$, and $R>1$); hence for each significantly overlapping Floquet eigenstate $|\psi_n\rangle$, the correction in Eq.~\eqref{eq:fourier_ft} is of order $\lesssim R^3T/\tau$, and hence much smaller than $1$. % is well-approximated by the first term in Eq.~\eqref{eq:fourier_ft}.

Since we chose $N$  much smaller than $\tau/T$,  for each $m$ between $0$ and $N$, $|\psi(mT)\rangle$ is still confined  within well $m\, (\mod q)$ of $\mathcal H_{\rm eff}(X,p)$.
Thus $|\psi_n\rangle$  can be written as % the form
\be 
|\psi_n\rangle = \sum_{k=1}^q|\chi^{k}_n\rangle + \mathcal O (R^3T/\tau),
\label{eq:chidef_0}
\ee
where $ |\chi_n^{k}\rangle$ consists of all  terms in  Eq.~\eqref{eq:fourier_ft}  where $m = k \, (\mod q)$ and hence only has support only in potential well $k$ of phase space.

The states $\{|\chi_n^k\rangle\}$ transform nontrivially under time translation by one period:
using that
 $\hat U(T)|\psi(mT)\rangle = |\psi([m+1]T)\rangle$, one can verify that % from the definition of $|\chi_n^k\rangle$ that 
\be 
\hat U (T) |\chi_n^{k}\rangle = e^{-i\varepsilon _n T}|\chi_n^{k+1}\rangle + \mathcal O (R^3T/\tau),
\label{eq:chi_transformation}
\ee
where $|\chi^{q+1}_n\rangle = |\chi^1_n\rangle$. % is identical to $k=1$.
Due to their separate regions of support in phase space,  the states $\{|\chi_n^k\rangle\}$ are mutually  orthogonal.
%The orthogonality implies that $\langle \psi_n|\psi_n\rangle = \sum_k \langle \chi_n^k|\chi_n^k\rangle + \mathcal O(R^3 T/\tau)$.
Since the right-hand side of Eq.~\eqref{eq:chidef_0} must have unit norm, the orthogonality of  the states $\{|\chi_n^k\rangle\}$ along with Eq.~\eqref{eq:chi_transformation}
%(which implies that they approximately have identical norms) 
require that $\langle \chi_n^k|\chi_n^k\rangle=1/q+ \mathcal O ( R^3T/\tau)$ for each $k$.

 \begin{figure}[t!]
    \includegraphics[width=0.99\columnwidth]{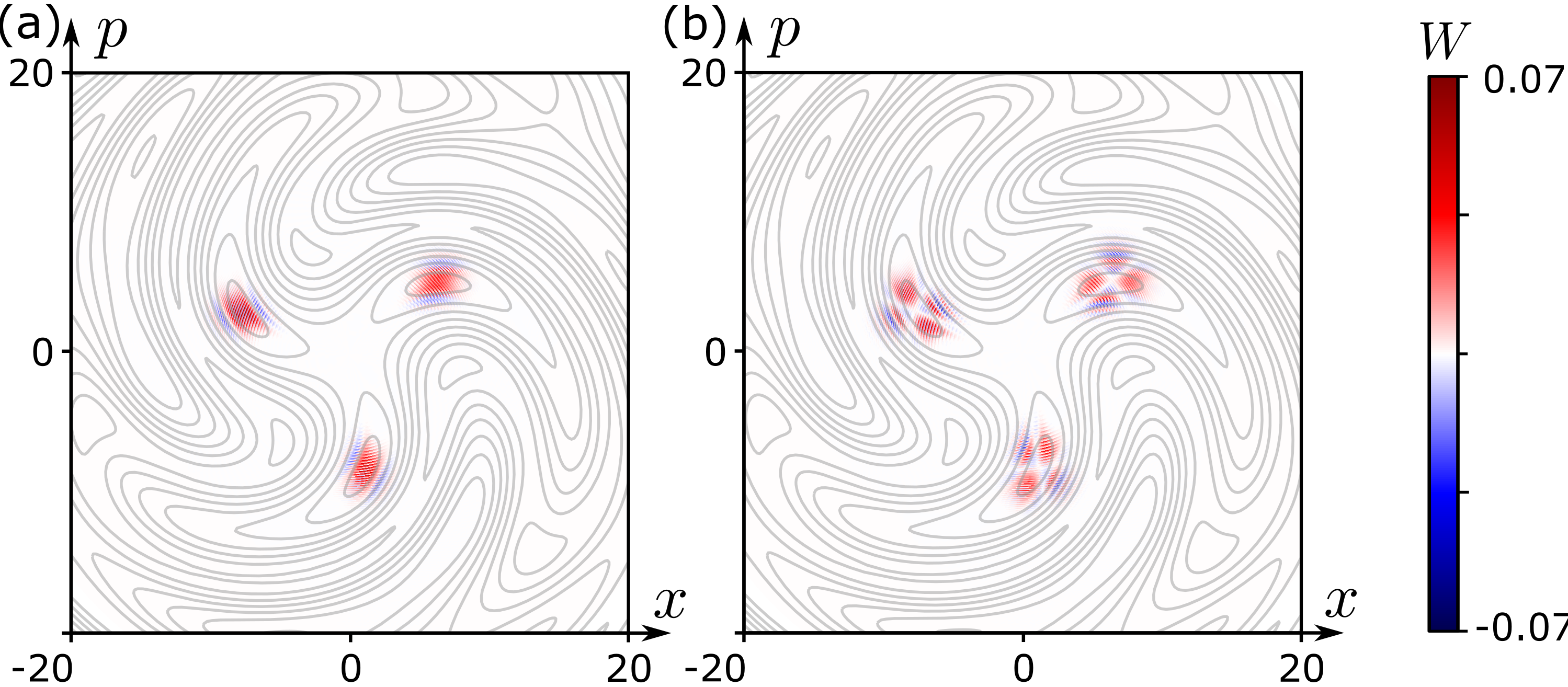}
    \caption{Wigner functions of two additional period-$3$ quasienergy-locked Floquet eigenstates for the same system as depicted in  Fig.~\ref{fig:contours}d, with support in  a well distinct from the eigenstate in Fig.~\ref{fig:contours}d. }
    \label{fig:other_wigners}
\end{figure}

We now use the states $\{|\chi^k_n\rangle\}$ to identify  a family of  Floquet eigenstates (including  $|\psi_n\rangle$), whose quasienergies are separated from each other by integer multiples of  $\Omega/q$ (up to corrections of order $1/\tau$) thus establishing the presence of quasienergy locking in the system.
To this end, we consider the states $|\bar\psi_n^1\rangle, \ldots|\bar\psi_n^q\rangle$, where 
$
|\bar\psi^{\ell}_n\rangle =\sum_{k=1}^q e^{-2\pi i \ell k /q} |\chi_n^{k}\rangle . %+ \mathcal O (R^3 T/\tau)
$
Using Eq.~\eqref{eq:chi_transformation}, we see that 
\be 
 U(T)|\bar\psi^\ell_n\rangle = e^{-i(\varepsilon_n+ \ell \Omega/q) T}|\bar\psi^\ell_n\rangle + \mathcal O(R^3 T/\tau ). 
\ee
Moreover, $|\bar\psi^\ell_n\rangle$ has norm $1+\mathcal O(R^3 T/\tau)$, due to the orthogonality and near-normalization of the states $\{|\chi_k^n\rangle\}$ [see text above Eq.~\eqref{eq:chi_transformation}].
Thus, up to a correction of order $R^3T/\tau$, $|\bar\psi^\ell_n\rangle$ is a normalized Floquet eigenstate of the system.
The existence of the approximate Floquet eigenstate $|\bar\psi_n^\ell\rangle$, along with the finite quasienergy level spacing, dictates~\cite{eigenstate_result} that there must exist  an {\it exact} Floquet eigenstate of the system, $|\psi_n^\ell\rangle\approx |\bar\psi_n^\ell\rangle$ whose quasienergy is approximately given by $\varepsilon_n+ \ell \Omega/q$.
Specifically, $|\psi_n^\ell\rangle$ can be written
\be
|\psi^{\ell}_n\rangle =\sum_{k=1}^q e^{-2\pi i \ell k /q} |\chi_n^{k}\rangle+ \mathcal O (R^5 T/\tau),
\label{eq:floquet_state_multiplet_final}
\ee
while the corresponding quasienergy is given by % the form    
\be
\varepsilon_n + \ell \Omega /q+ \mathcal O (R^3 T/\tau).\label{eq:triplet}
\ee
For each Floquet eigenstate $|\psi_n\rangle$ that significantly overlaps with the state $|\psi(0)\rangle$ (according to the definition above), this construction can be made for each $\ell =1 ,\ldots q$ (note that $|\psi_n^q \rangle = |\psi_n\rangle$); hence, each Floquet eigenstate with significant support in the potential wells of $\mathcal H_{\rm eff}$ must form a part of a multiplet of $q$ Floquet  states with the properties outlined in Eqs.~\eqref{eq:qel}-\eqref{eq:chi_relation} from the main text. 
%above. 
This establishes the presence of  quasienergy locking of the qubit-cavity system in the   regimes, and concludes this Appendix.

\section{Floquet eigenstates with support in distinct potential wells}
\label{app:other_wigners}

In this Appendix, we show Wigner functions for additional period-$3$ quasienergy-locked Floquet eigenstates of  the qubit-cavity model for the parameters considered in Fig.~\ref{fig:contours}bd (see main text for further details). 
These plots demonstrate that the distinct potential wells of $\mathcal H_{\rm eff}(x,p)$ for these parameters (see Fig~\ref{fig:contours}b) support distinct triplets of quasienergy-locked Floquet eigenstates, and moreover that the  same  well may support several triplets. 

In Fig.~\ref{fig:other_wigners}, we depict the Wigner functions for two quasienergy-locked Floquet eigenstates of the system distinct from the one depicted in Fig.~\ref{fig:contours}d.
Each of the  Floquet eigenstates depicted in Fig.~\ref{fig:other_wigners} form a part of its own triplet of quasienergy-locked Floquet eigenstates which, respectively, have (nearly) identical Wigner functions to the Wigner functions depicted in Fig.~\ref{fig:other_wigners}. 

Note that the two Floquet eigenstates shown in Fig.~\ref{fig:other_wigners} have support in the same well, and that this well is distinct from the one where the eigenstate shown in Fig.~\ref{fig:contours}d has support.
This confirms that the distinct potential wells of $\mathcal H_{\rm eff}(x,p)$ support distinct triplets  of quasienergy locked Floquet eigenstates, and that the same well may support multiple triplets. 

We finally note that the nodal lines of the Wigner functions in Fig.~\ref{fig:other_wigners} coincide very closely with the contours of $\mathcal H_{\rm eff}(x,p)$  (gray lines), hence  further supporting the discussion in Sec.~\ref{sec:effective_hamiltonian}.

\end{document}